%% file: main.tex
\documentclass[sigconf]{acmart}
\usepackage{tikz}

\usepackage{amsmath,amsfonts}
\usepackage{algorithmic}
\usepackage{graphicx}
\usepackage{subcaption}
\usepackage{textcomp}
\usepackage{xcolor}
\usepackage{float}
\usepackage[capitalise]{cleveref}

\usepackage{algorithmic}
\usepackage{algorithm}
\usepackage{booktabs}
\usepackage{adjustbox}
\usepackage{xspace}

\usepackage{enumerate}

\usepackage[avoid-all]{widows-and-orphans}

\AtBeginDocument{%
  \providecommand\BibTeX{{%
    \normalfont B\kern-0.5em{\scshape i\kern-0.25em b}\kern-0.8em\TeX}}}

\newcommand{\name}{\textit{SkinSense}\xspace}

\begin{document}

\title[\textit{SkinSense}]{\textit{SkinSense:} Efficient Vibration-based Communications Over Human Body Using Motion Sensors}

\author{Raveen Wijewickrama$^*$}
\affiliation{
\institution{University of Texas at San Antonio}
\country{USA}
}
\email{raveen.wijewickrama@utsa.edu}

\author{Sameer Anis Dohadwalla$^*$}
\affiliation{
\institution{Indian Institute Of Technology Dharwad}
\country{India}
}
\email{180020033@iitdh.ac.in}

\author{Anindya Maiti}
\affiliation{
\institution{University of Oklahoma}
\country{USA}
}
\email{am@ou.edu}

\author{Murtuza Jadliwala}
\affiliation{
\institution{University of Texas at San Antonio}
\country{USA}
}
\email{murtuza.jadliwala@utsa.edu}

\author{Sashank Narain}
\affiliation{
\institution{University of Massachusetts Lowell}
\country{USA}
}
\email{sashank_narain@uml.edu}

\thanks{$^*$ These authors contributed equally to this work.}
\renewcommand{\shortauthors}{Wijewickrama et al.}

\begin{abstract}
\input{abstract}

\end{abstract}


\renewcommand\footnotetextcopyrightpermission[1]{} %
\pagestyle{plain} %

\maketitle

\input{introduction}

\input{background}

\input{system}

\input{evaluation}

\input{discussion}

\input{related}

\input{conclusion}

\bibliographystyle{ACM-Reference-Format}
\bibliography{references}

\end{document}

%% file: abstract.tex
Recent growth in popularity of mobile and wearable devices has re-ignited the need for reliable and stealthy communication side-channels to enable applications such as secret/PIN sharing, co-location proofs and user authentication. Existing short-range wireless radio technology such as Bluetooth/BLE and NFC, although mature and robust, is prone to eavesdropping, jamming and/or interference, and is not very useful as a covert communication side-channel. This paper designs and implements \name, a vibration-based communication protocol which uses human body/skin as a communication medium to create a low-bandwidth and covert communication channel between user-held mobile and wearable devices. \name employs a novel frequency modulation technique for encoding bits as vibration pulses and a spectrogram-based approach to decode the sensed motion data (corresponding to the encoded vibration pulses) to reconstruct the transmitted bits. \name is comprehensively evaluated for a variety of operational parameters, hardware setups and communication settings by means of data collected from human subject participants. Results from these empirical evaluations demonstrate that \name is able to achieve a stable bandwidth of upto 6.6 \emph{bps}, with bit error rates below 0.1 in our custom hardware setup, and can be employed as a practical communication side-channel.

%% file: introduction.tex
\section{Introduction} %
\label{sec:introduction}

The ubiquity of smart wearables has rapidly increased in the past few years, wherein inexpensive devices such as smartwatches, fitness bands, smart-glasses, and smart-shoes provide users with a variety of additional capabilities beyond that of a smartphone.
Apart from a small subset of these smart wearables which can operate autonomously, the vast majority of smart wearables are used in conjunction with a paired user-held mobile device such as a smartphone. 
Currently, this communication between the mobile device (smartphone) and paired smart wearables is primarily accomplished using short-range wireless radio technology such as Bluetooth/BLE, NFC, and Wi-Fi. Although these technologies are very mature and enable robust communication between mobile devices, wireless radio signals are prone to eavesdropping, jamming and/or interference, resulting in loss of communication or privacy or both. 
There is a genuine need for reliable and stealthy non-radio communication side-channels in several use-cases and applications involving these devices. For instance, a low bandwidth, non-radio based protocol could be used to swiftly and securely send a short secret code or a PIN from a smartphone to a smart wrist wearable to automatically initiate and establish a high-bandwidth and secure radio channel, such as Bluetooth or WiFi. Besides secret sharing, the availability of such an out-of-band communication side-channel could prove to be convenient and practical for several other security applications such as proof of device co-location and user authentication.
Some recent proposals on alternate body-area communication channels have suggested the use of optical \cite{chevalier2015wireless,chevalier2015optical} and acoustic signals \cite{laput2016viband,galluccio2012challenges} to enable low-bandwidth message transfers between mobile devices. Such communication channels, however, are not robust and often perform poorly in the presence of noise or non-ideal ambient conditions, and are not very covert (i.e., easily detectable) \cite{rushanan2014sok}.

Other proposals in this direction have employed vibration motors, often embedded in mobiles devices to provide haptic feedback to users, as a transmitter  \cite{hwang2012privacy,roy2015ripple,roy2016ripple,ma2020skin}. In such techniques, fine-grained motion sensors such as accelerometers and gyroscopes (also found on most modern mobile devices) have been adopted as receivers \cite{michalevsky2014gyrophone}. 
Hwan et al. \cite{hwang2012privacy} were one of the first to propose the pairing of a vibration motor and an accelerometer to establish a protocol for low-bandwidth communication between two smartphones kept on the same (hard) surface. Roy et al. \cite{roy2015ripple} followed up by proposing a scheme, called Ripple, which provided improved bandwidth and added security. 
Kim et al. \cite{kim2015vibration} proposed the use of a smartphone as a vibration transmitter for a custom medical device with an accelerometer implanted under the skin acting as the receiver.
Later, Sen and Kotz \cite{sen2020vibering} proposed the use of a smart ring as a vibration transmitter to communicate with Internet-of-Things (IoT) devices equipped with accelerometers, however for their scheme to work, the transmitter (smart ring) needed to be in direct contact with the receiver (IoT device) for reliable communication. 

In this work, we take inspiration from these earlier research efforts to explore the feasibility of employing vibration motors and accelerometers to build a reliable communication protocol between two user-held mobile devices (e.g., a smartphone and a wrist wearable), by using the human body/skin as the underlying physical medium for vibration signal propagation. Given that modern mobile devices such as smartphones and wearables are always in contact with the human body/skin, and that they come pre-equipped with vibration motors and motion sensors, such a communication technique could be naturally deployed without requiring any special setup/hardware.
Also, as the user's own body/skin acts as a communication channel, the scheme would be naturally covert and robust against trivial eavesdropping. 

Although vibration-based communication techniques have been proposed before, additional research and experimentation is required to build a robust (against noise and movements) and easy-to-deploy communication system that employs vibrations traveling through human skin/body, which is what this work attempts to accomplish. To this end, the main challenges include identifying how vibrations can be modulated so that they can effectively propagate via human skin/body, and efficiently using an accelerometer to sense these vibrations in the presence of volatile body movements. Moreover, inherent anatomical and biomechanical differences between individuals can also affect the efficiency of the communication channel. Consequently, the main requirements as we design such a communication technique are: (a) \textbf{robustness} to vibration noise and anatomical differences that are inherently present in human body/skin, (b) a sufficient level of \textbf{data rate} suitable for low-bandwidth communications, and
(c) reliable communication under \textbf{mobility} such as while traveling in a vehicle. 

Our key contribution in this paper is the design, implementation and evaluation of a novel vibration-based communication protocol, called \name, which utilizes human body/skin as a communication medium in order to create a low-bandwidth and covert communication channel between user-held mobile and wearable devices.
We demonstrate the feasibility and performance of \name by employing both a custom hardware setup, which enables the flexibility of being able to manipulate the different hardware and protocol parameters for a comprehensive evaluation, and consumer-level devices.
By collecting data from a diverse set of human subject participants, we comprehensively evaluate \name under different/varying operational parameters and communication settings, including, distance between transmitter and receiver, device orientation, ambient noise, sampling rate, communication direction and motion sensing hardware (accelerometer versus gyroscope). In addition to this, we also conduct an in-depth analysis of the impact of the various \name design parameters on communication performance and error rates, empirically evaluate its vulnerability to acoustic or Sound of Vibration (SoV) attacks, and study its energy requirements. Lastly, we systematically compare the performance of \name with other efforts in the literature and conduct a small user-study to gauge the perception and preferences of end users using such a communication side-channel.

%% file: background.tex
\section{Background} %
\label{sec:background} 

We now briefly introduce some background information relevant to our proposed communication protocol.

\begin{figure}[htbp]
         \begin{subfigure}[b]{0.36\linewidth}
             \centering
             \includegraphics[width=\textwidth]{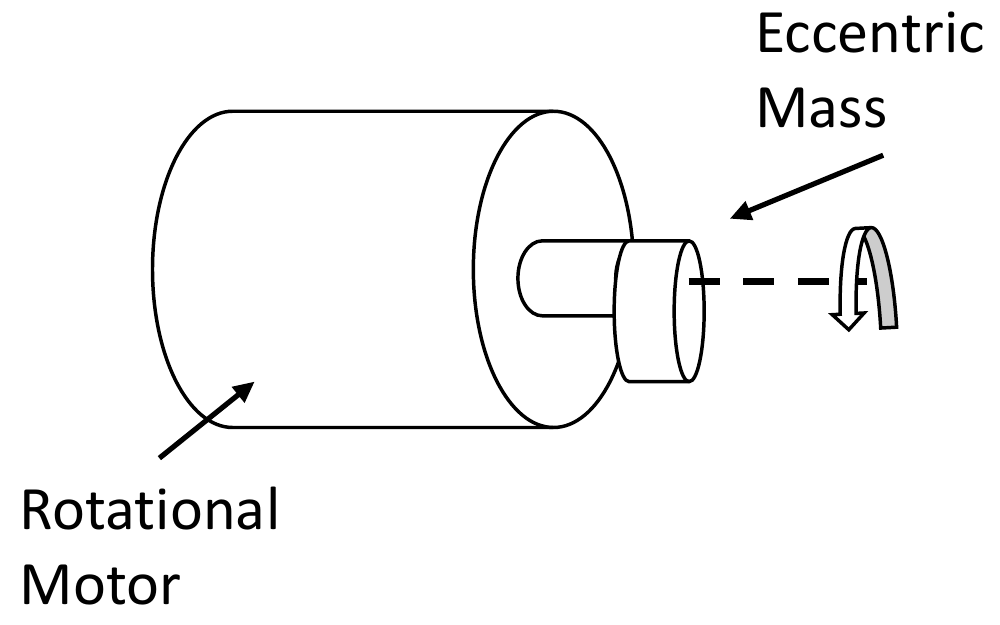}
             \caption{}
             \label{fig:erm}
         \end{subfigure}
         \quad
         \begin{subfigure}[b]{0.56\linewidth}
             \centering
             \includegraphics[width=\textwidth]{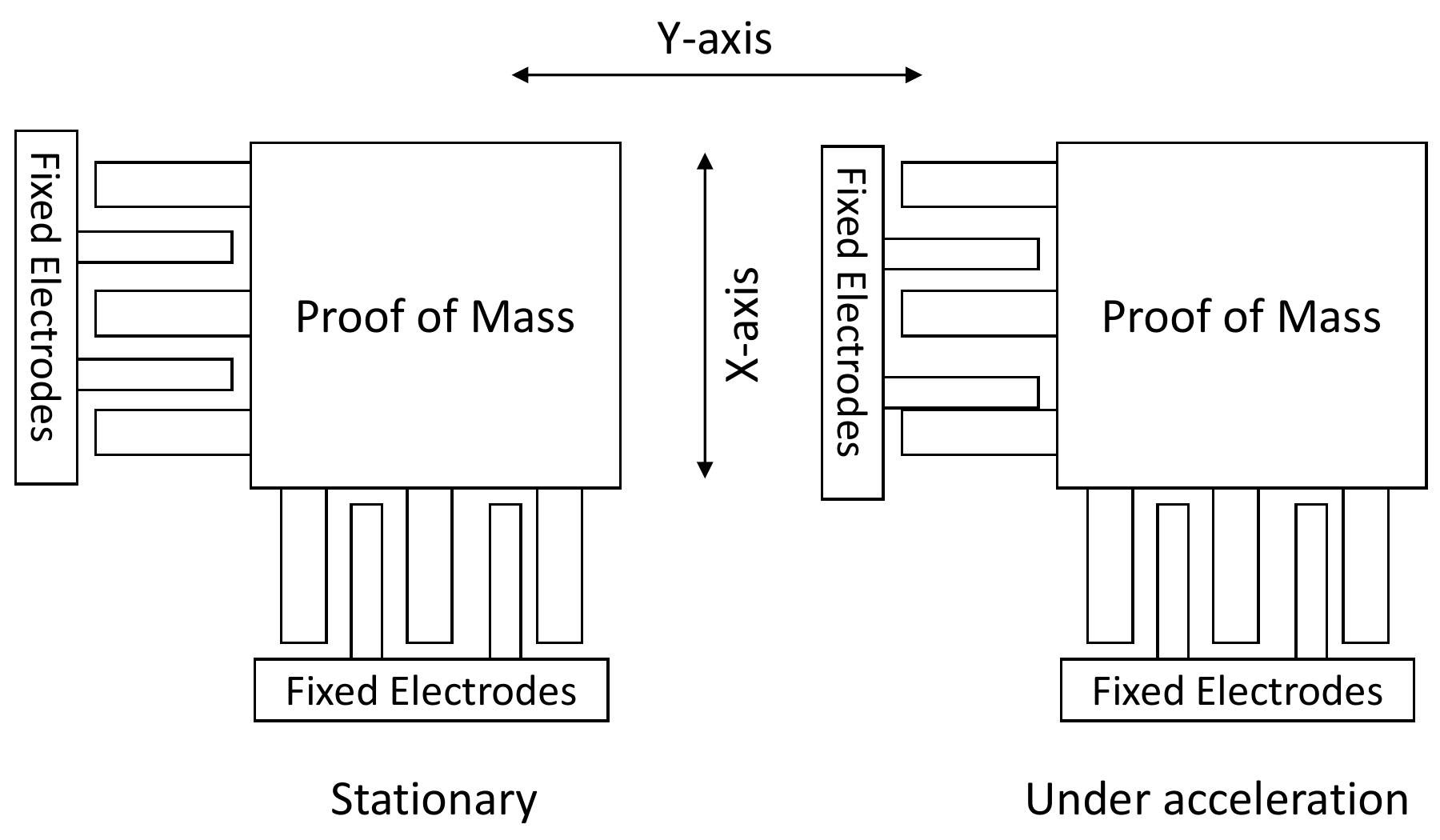}
             \caption{}
             \label{fig:archi_accel}
         \end{subfigure}
             \caption{Architecture of (a) ERM motor and (b) capacitive accelerometer.}
            \label{fig:motors}
\end{figure}

\subsection{Vibration Motor} %
\label{sub:vibration_motor}
A vibration motor is a mechanical device that uses motion from an unbalanced metallic mass to produce oscillations or vibrations \cite{Gao2021}. They are usually found inside most modern %
mobile devices such as smartphones and smartwatches to provide haptic feedback or notifications to users. There are two main types of vibration motors: (i) \emph{Eccentric Rotating Mass (ERM)} motors, and (ii) \emph{Linear Resonant Actuator (LRA)} motors. ERM motors are powered by direct current (DC) and employ an asymmetric mass (\cref{fig:erm}) which moves eccentrically when rotated \cite{seim2015perception}. The amplitude and frequency of the vibrations can be controlled by varying the input DC voltage, and they both increase with higher voltages \cite{ali2021fine,roy2015ripple}. LRA motors, on the other hand, consist of a magnetic coil that pushes a mass up and down to create vibrations, enhanced by a spring. LRA is driven by a precise amount of AC current so as to achieve resonant frequency of the spring, which limits the vibration motor's amplitude and frequency within a very narrow band.
Huang et al. \cite{huang2016experiment} showed that ERM motors are better suited for applications requiring complex encoding of the vibration signal, compared to LRA motors which work well only for simple binary encoding. Therefore, we employ an ERM-based DC coreless motor in this work.

\textbf{Motor Control.} %
\emph{Pulse Width modulation (or PWM)}, a technique to reduce the average power delivered to a load by effectively breaking the input electrical signal into discrete parts, is often used to control intertial loads such as vibration motors found in most mobile devices \cite{barr2001pulse, luo2018undersampled}. The average value of voltage (and current) supplied to a motor in PWM is regulated by turning the switch between the power supply and the motor ON and OFF at a fast and variable rate. In other words, PWM can control the operation of a motor by providing it with a series of electrical pulses or ``ON-OFF'' signals. 
The power to the motor is controlled by varying the width of these pulses, which in turn, varies the average DC voltage applied to the motor. 
Ultimately, PWM enables greater control over the motor without altering the voltage level of the supply voltage, which is ideal for encoding signals over motor vibrations. We utilize a selected range of PWMs to control the motor based on the feedback observed via vibration signals captured using motion sensors, as discussed later in \cref{sub:algorithms}.

\subsection{Accelerometer} %
\label{sub:accelerometer}

Accelerometers are used to measure acceleration or speed changes of a device with respect to the surrounding environment. Most modern mobile and wearable devices are equipped with \emph{micro-electromagnetic (or MEMS)} \cite{d2019review,varanis2018mems} accelerometers, which enable applications ranging from simple step counting to complex activity recognition \cite{shoaib2016complex}. 
There are two main types of MEMS accelerometers in use today: (i) \emph{piezoresistive} accelerometers, and \emph{capacitive} accelerometers \cite{gao2004micromachined}. Piezoresistive accelerometers consist of a mass attached to a piezoelectric crystal. When vibrations or movements occur on the accelerometer, while the mass itself remains unchanged, it makes the crystal either compress or stretch depending on the direction and magnitude of acceleration. Capacitive accelerometers (\cref{fig:archi_accel}) consist of a proof-of-mass suspended between two plates, one which is fixed and other which is free to move inside the accelerometer housing. When vibrations/movements occur on the accelerometer, the distance between the plates change proportionally to the acceleration, resulting in a change of capacitance \cite{VENKATANARAYANAN201447}.
Most modern devices are equipped with capacitive accelerometers due to their smaller size and ability to measure low-frequency motion \cite{d2019review}.
Accelerometers produce raw measurements along three axes, with gravity components applied to whichever axis is pointing to the ground. 
Another type of motion sensor commonly found in consumer mobile devices is the gyroscope, which in contrast to accelerometers, measures a device's angular velocity. Later in \cref{sec:evaluation}, we comparatively evaluate our proposed communication protocol using both the accelerometer and gyroscope sensors.

\subsection{Mechanics of Vibrations} %
\label{sub:mechanics_vib}

\begin{figure}[h]
\centering
\includegraphics[width=0.6\linewidth]{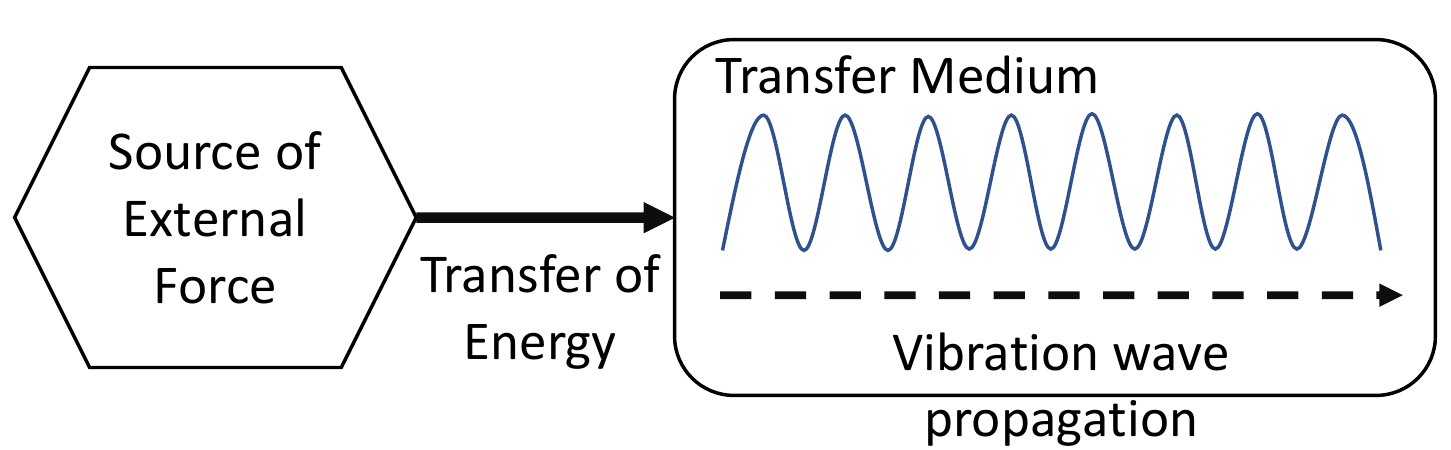}
\caption{Vibrational energy propagation from a source to a medium.}
\label{fig:wave_propagation}
\end{figure}

Vibrations are a form of mechanical oscillations, defined as repetitive back and forth movement of an object between two states or positions about an equilibrium point \cite{rittweger2020manual}. There are 2 kinds of vibrations, \emph{free} and \emph{forced}. Free vibrations occur as a result of a brief energy transfer from an external force on to an object (e.g., one time picking of a guitar string) \cite{rittweger2020manual}. In contrast, forced vibration is a continuous periodic external force applied to an object (e.g., repeatedly pushing a playground swing).
Vibrations can be fully described using two parameters, namely, \emph{frequency} and \emph{amplitude}. 
Frequency of vibration describes how fast the vibrating object is moving, while amplitude is the maximum displacement of the vibrating object in motion, i.e., the strength of the vibrations.
When an object vibrates, its particles are disturbed by the vibrational energy propagating from one point of the object to another, also called the \emph{vibration wave} (\cref{fig:wave_propagation}). 
How well these waves propagate through the object (or medium) depends on its molecular structure (such as density). If an object has tightly packed molecules, it is known to have a higher rigidity (e.g., wood or steel). Such objects or mediums are better at transmitting vibration waves making them travel farther, faster and longer compared to less rigid mediums (e.g., a cushion pillow) \cite{blake1961basic}.
Vibration waves have a characteristic wavelength (\(\gamma\)), which is the distance between two adjacent crests (highest points) of the wave. When a wave travels from one medium to another, although the wavelength changes, the frequency remains unchanged \cite{rittweger2020manual}. This is defined as \(v = f  \gamma\), where \(v\) is the velocity and \(f\) is the frequency of the wave. Thus, the velocity and wavelength of the vibration waves originating from some source and transferring to an object/medium could change depending on the rigidity of the object.

In this work, we are attempting to leverage vibrations originating from a handheld device such as a smartphone to design a communication mechanism, where the vibration waves %
propagate to the receiver through the skin of the hand (holding the handheld).
Previous works on vibration-based communication schemes have employed objects such as wooden tables as the transmission medium \cite{roy2015ripple}. These efforts have been able to achieve very high data rates and accuracy, primarily because wood is a good propagation medium for vibration waves due to its high rigidity. In contrast, human skin absorbs more vibrations due to its lower rigidity, making it less desirable as a transmission medium for vibration waves \cite{ali2021fine}.
Consequently, designing a vibration-based communication scheme relying on the propagation of the vibration waves through the human skin is a much more challenging endeavor.
\emph{Prior studies have shown that the propagation of vibration waves via the human skin is correlated to the vibration frequency of the source, and that the vibration decay is quicker at low and high frequencies (i.e., vibrations cannot propagate for longer distances) compared to intermediate frequencies of the source \cite{shah2019vibration,manfredi2012effect}.}
Since our work considers a communication scenario of a handheld device and a wrist wearable, which naturally results in a certain amount of distance (with distances ranging from 15 \emph{cm} to 2 \emph{cm} depending on the motor placement on the handheld device) between the two devices when worn on the hand, understanding and analyzing this vibration decay and the effect of vibration frequency was important in designing an effective communication framework. In order to identify these intermediate frequencies which would help in designing \name, we do a \emph{sine sweep test} as explained further in \cref{ssub:transmitter_design}.

Based on the high-level discussion so far on the mechanics of vibrations, we now formally describe how vibrations created by motors (e.g., ERM motors) can be used to send vibration waves (or signals) through human skin as a medium.
The centrifugal force generated by an ERM motor is given by:
\[F = m  r  \omega^2\]
where \(F\) is the force (in $Newtons$), \(m\) is the mass of the eccentric mass ($kg$), \(r\) is the radius of the eccentric mass ($meters$) and \(\omega\) is the angular velocity (speed of the motor) in $rads/sec$ \cite{rittweger2020manual}.
As \(m\) and \(r\) are physical properties of the motor that cannot be changed, the centrifugal force generated by the motor can only be changed by manipulating the angular velocity, \(\omega\). 
The vibration frequency and amplitude in an ERM motor cannot be changed independently and they both increase linearly based on the voltage provided. 
ERM motor speeds are proportional to the applied voltage, therefore amplitude/frequency changes can be done by manipulating the voltage via pulse width modulation (PWM) as discussed above in \cref{sub:vibration_motor}. 
As clarified earlier, identifying a set of intermediate frequencies that propagates well from a handheld device to a wrist wearable via human skin as the medium is the first step in designing an effective communication scheme via skin. To modulate a set of frequencies, we rely on a PWM based technique and then use these different frequencies to encode data in \name as described in detail in \cref{sec:system}.

%% file: system.tex
\section{System} %
\label{sec:system}

We now present the design and other technical details of \name by first providing an overview of the system architecture, followed by design details of the communication protocol. We then present the technical details of the encoding (modulation) and decoding (demodulation) algorithms followed by an outline of different hardware setups that we employ in the implementation of \name. Finally, we present details of the human subject data and performance metrics used in its evaluation.

\subsection{System Overview} %
\label{sub:system_overview}

\begin{figure}[H]
\centering
\includegraphics[width=\linewidth]{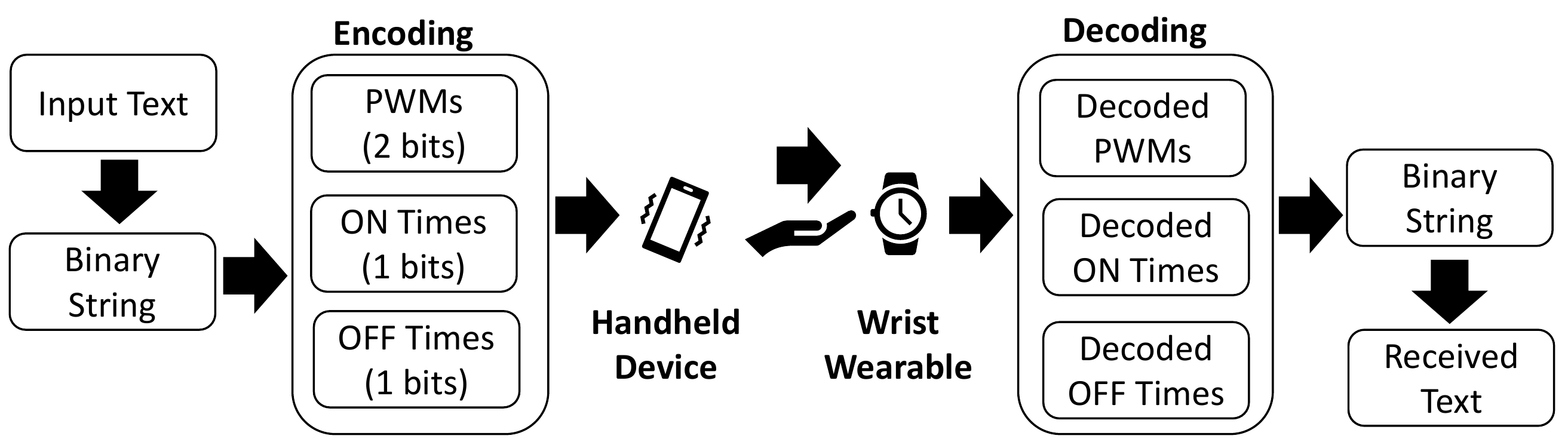}
\caption{\name communication protocol.}
\label{fig:overview}
\end{figure}

\cref{fig:overview} shows a high-level overview of \name enabling vibration-based communication between a handheld device and a wrist wearable by using the user's hand (specifically, the skin tissues) as the communication channel. We consider a half-duplex communication channel, i.e., the handheld device and wrist wearable can both act as a transmitter and receiver, however information can flow in only one direction at a time. Moreover, we consider that the communicating devices are located on the same hand, but they do not need to be in physical contact with each other. 
The message that needs to be transmitted from the handheld device to the wrist wearable (or vice versa) is first encoded into a stream of vibration signals by using an encoding (or modulation) algorithm described in \cref{subsub:enc_algorithm}. The vibration motor on the transmitting device then emits these vibration signals, which is carried via the skin tissues of the hand to the receiving device. The receiving device's motion sensor records these vibrations and decodes the encoded message using a decoding algorithm outlined in \cref{subsub:dec_algorithm}.

\subsection{Transmitter Design} %
\label{ssub:transmitter_design}
Our final transmission algorithm was arrived at after many rounds of iterations and preliminary investigations where we scrutinized on how vibrations travel through the human skin and its ability to be captured by an accelerometer.
As a first step, our goal was to examine the %
possibility of identifying/differentiating multiple vibration frequencies for data encoding, which would effectively allow us to significantly increase the bandwidth of our communication protocol (compared to using only a single frequency encoding scheme). %
Specifically, we analyzed how vibrations of different frequencies, after being generated by the motor and traveling via the skin, gets captured on an accelerometer. 
Once specific operable frequencies with clear separations were identified, our next goal was to minimize the vibration times, i.e. the time of a single vibration pulse (\texttt{ON} times) and the interval between two vibration pulses (\texttt{OFF} times). At the same time, to further increase the bandwidth and data rate, %
we also found that time domain data encoding could also be used in conjunction to frequency-based encoding by %
reliably determining vibration \texttt{ON} times and \texttt{OFF} times.
The specific details of selecting operating frequencies and time modulation are presented next.

\subsubsection{Identifying the operable frequencies}

In order to identify and characterize the range of frequency bands to operate the transmitter motor in, such that the receiver accelerometer would have a distinct response to the vibration signal, we did a \emph{sine sweep test} \cite{roy2015ripple} by test operating the motor in the range of 20 (low) to 240 (high) PWM values and analyzed the corresponding accelerometer signal at the receiver. 

\begin{figure}[h]
\centering
\includegraphics[width=0.7\linewidth]{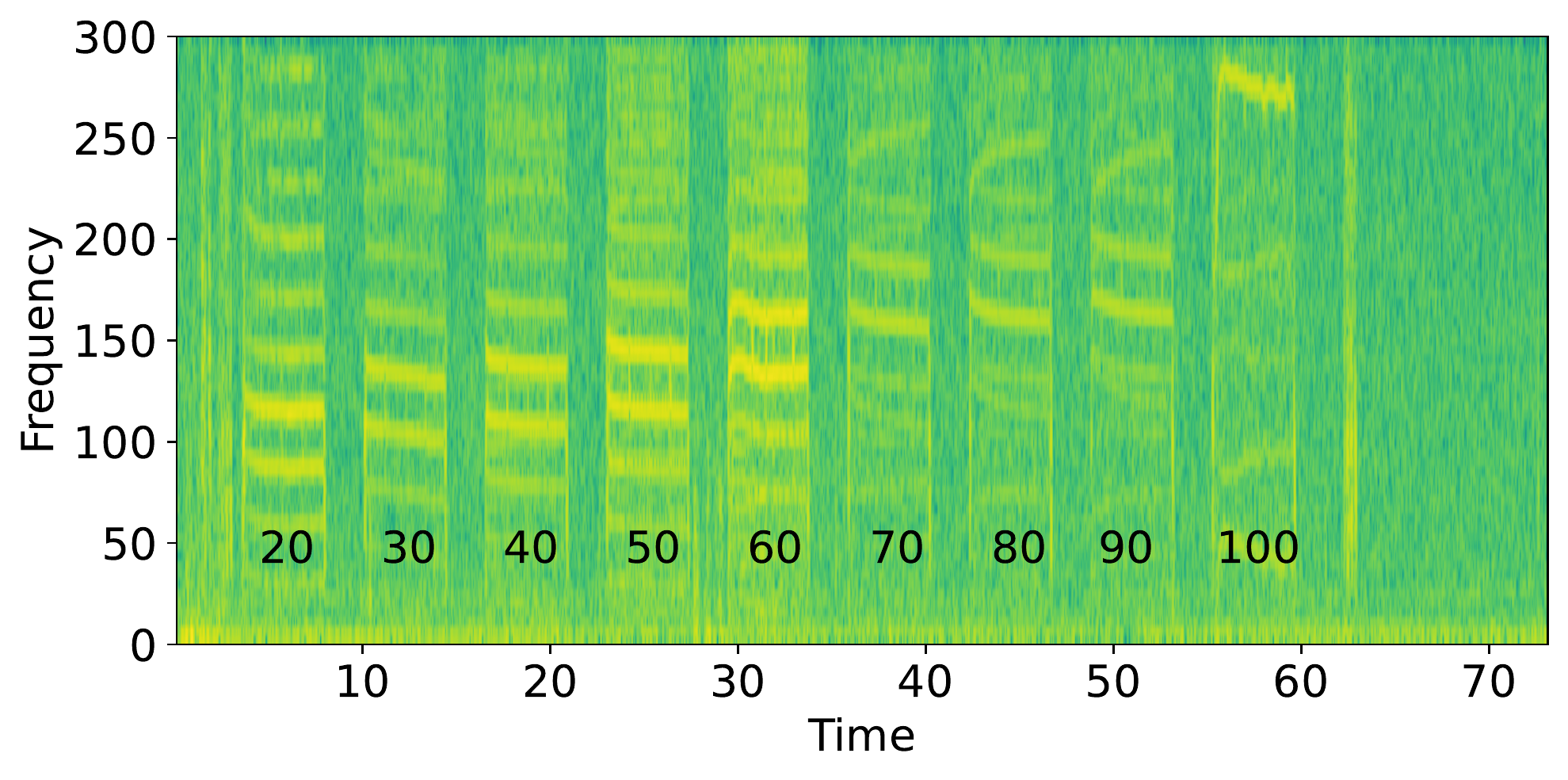}
\caption{Spectrogram of \texttt{PWM}s ranging from 20 to 100.}
\label{fig:spectrogramPWMs}
\end{figure}

We observed that the frequency response captured by the accelerometer peaks around \texttt{PWM} 60 and fades away starting around \texttt{PWM} 100 (refer \cref{fig:spectrogramPWMs}). 
Based on this observation, we determined that the motor reaches its resonant frequency at around \texttt{PWM} 60. While a simple \texttt{ON} or \texttt{OFF} binary encoding mechanism would only require one operational frequency for the motor, this will limit the amount of data that can be encoded using vibrations.
After identifying the operable frequencies to be in the range of \texttt{PWM} 20-100, we then closely analyzed which specific \texttt{PWM} frequencies to work with. For the frequency analysis, we closely observed the frequency response of \texttt{PWMs} from 20 up to 100 in steps of 10.
We subsequently found \texttt{PWMs} 20, 30, 60 and 100 to be the ones with minimal interference with neighboring \texttt{PWMs}, and thus employ them as the final set of \texttt{PWM} parameters in \name implementation.

\subsubsection{Time based modulation}

We next studied the possibility of minimizing the \texttt{ON} time and \texttt{OFF} time of the vibrations to further increase the data rate. For that, we analyzed how accurately we can infer the \texttt{ON} time and \texttt{OFF} time windows of vibrations using the accelerometer signal for a range of \texttt{ON} and \texttt{OFF} time values.
These \texttt{ON} and \texttt{OFF} times are affected by a phenomenon called the \emph{ringing effect} \cite{roy2015ripple}, where the vibration may remain in the medium for sometime before completely dying down when the voltage is cut off to the motor. When choosing \texttt{OFF} times, we need to specifically select values that accommodate the remnants of the previous vibration pulse such that, the next vibration pulse is not affected by it to prevent inaccuracies during demodulation.
We tested \texttt{OFF} times in the range of 150-1000 \emph{ms} and were able to clearly distinguish between \texttt{OFF} time values of 250 \emph{ms}, 500 \emph{ms}, 750 \emph{ms} and 1000 \emph{ms}. Any values lower than 150 {ms} proved to be too short, leading to interference between consecutive vibration pulses.
Therefore, without affecting the overall accuracy, we select 150 \emph{ms} and 300 \emph{ms} as the \texttt{OFF} time window values in our algorithm implementation.
After finalizing appropriate \texttt{OFF} time values, we determine \texttt{ON} time window values, i.e. the amount of time the vibration motor stays on in a single window. Similar to the \texttt{OFF} time analysis, we tested \texttt{ON} times in the range of 250-1000 \emph{ms}, followed by \texttt{ON} time values below 250 \emph{ms}, however values below 250 \emph{ms} proved to be too small to be accurately identified in the accelerometer signal. Thus, without affecting the overall accuracy, we fixed 250 \emph{ms} and 500 \emph{ms} as the \texttt{ON} time window values for our implementation. 
The full hardware setup used for this analysis is detailed in \cref{subsub:custom_hardware}. An independent analysis may be required to determine the optimal parameters for each motor-accelerometer pair since different motor types may have varying resonant frequencies, and different accelerometer types may have varying sensitivities.

\subsection{Communication Algorithms} %
\label{sub:algorithms}

\name employs a \texttt{PWM} based frequency modulation technique for encoding bits as vibration pulses 
and a spectrogram based approach to decode the sensed motion data (corresponding to the vibration pulses) to reconstruct the transmitted bits, as outlined next.

\subsubsection{Encoding Algorithm}
\label{subsub:enc_algorithm}
The transmission algorithm takes as input a sequence of bits to be transmitted and outputs a sequence of parameters (\texttt{PWMs}, \texttt{ON} times and \texttt{OFF} times) to be passed on to the vibration motor. The \texttt{PWMs} are a measure of the voltage given to the motor, which in turn controls its frequency. The \texttt{ON} times signify the time in milliseconds (\emph{ms}) for which the vibration motor is switched on, while the \texttt{OFF} times signify the time in milliseconds between two consecutive vibration pulses.
By modulating these three parameters independently, we are able to transmit the bit string as a series of vibrations of differing times and frequencies. 

\name currently uses four values of \texttt{PWMs} (20, 30, 60, 100), two values of \texttt{ON} times (250 \emph{ms}, 500 \emph{ms}) and two values of \texttt{OFF} times (150 \emph{ms}, 300 \emph{ms}) for modulation, as described earlier in \cref{ssub:transmitter_design}. As a result, we are able to encode $log\sb{2}4=2$ bits using the \texttt{PWM} values and $log\sb{2}2=1$ bit using the \texttt{ON} time and \texttt{OFF} time values, respectively. 
In order to transmit a bit sequence using the above setup and set of parameters, we need to partition it into 4-bit long words. Then, we encode each 4-bit word as follows: The first two bits of the word are encoded using one of the four \texttt{PWM} values, the third bit using one of the two \texttt{ON} time values and the fourth bit using one of the two \texttt{OFF} time values. 
At the end of transmitting message, we append a pilot sequence to improve accuracy while decoding. 

\subsubsection{Decoding Algorithm}
\label{subsub:dec_algorithm}
The decoding algorithm takes as input the raw accelerometer values sensed at the receiver and outputs the decoded bit sequence. %
The sequence of steps involved in the decoding algorithm are as follows:
\begin{enumerate}[\hspace{0pt}(1)]
    \item Spectrogram computation: Raw signal filtering and normalization.
    \item Peak detection: Deriving the encoded parameters.
    \item Symbol separation: Symbol separation to identify the encoded \texttt{PWM} values, \texttt{ON} time values and \texttt{OFF} time values.
    \item Pilot sequence based mapping and decoding: Decode the message by generating the corresponding bits associated with each \texttt{PWM} values, \texttt{ON} time values and \texttt{OFF} time values.
\end{enumerate}

\textbf{Spectrogram.}
The spectrogram of raw accelerometer time-series data samples are first computed in the decoding process. Essentially, a spectrogram is a matrix which depicts the strength of the accelerometer signal over time at different frequencies (a spectrum of frequencies), i.e., each position of the matrix corresponds to a point in frequency and time.
In other words, the time-series data is converted from the time domain to the frequency domain (using FFT with a window size of 128 samples and an overlap of 124 samples). 
In order to reduce the effect of noise, we filter out all frequencies below 40 {\emph{Hz} from the spectrogram. These frequencies fall below our band of operation and correspond mainly to low frequency noise which may be present in the human hand.
\cref{fig:spec_before_after} shows the frequency spectrum before and after removing this noise. We then perform mean and variance normalization on the resulting spectrogram in order to eliminate any constant biases that may be present.

\begin{figure}[htbp]
\begin{subfigure}[b]{0.32\linewidth}
 \centering
 \includegraphics[width=\textwidth]{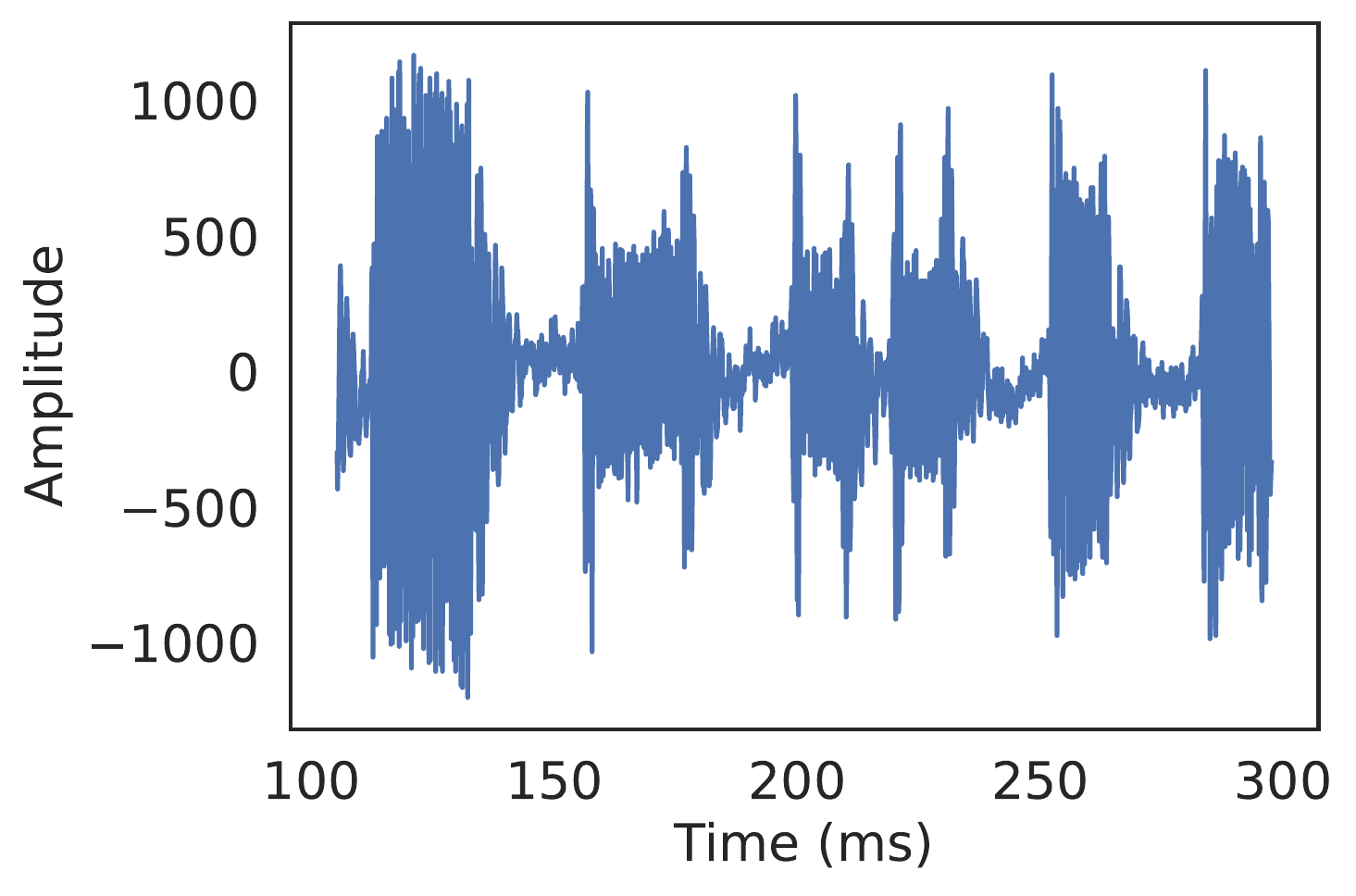}
 \caption{}
 \label{fig:time_domain}
\end{subfigure}
\begin{subfigure}[b]{0.32\linewidth}
 \centering
 \includegraphics[width=\textwidth]{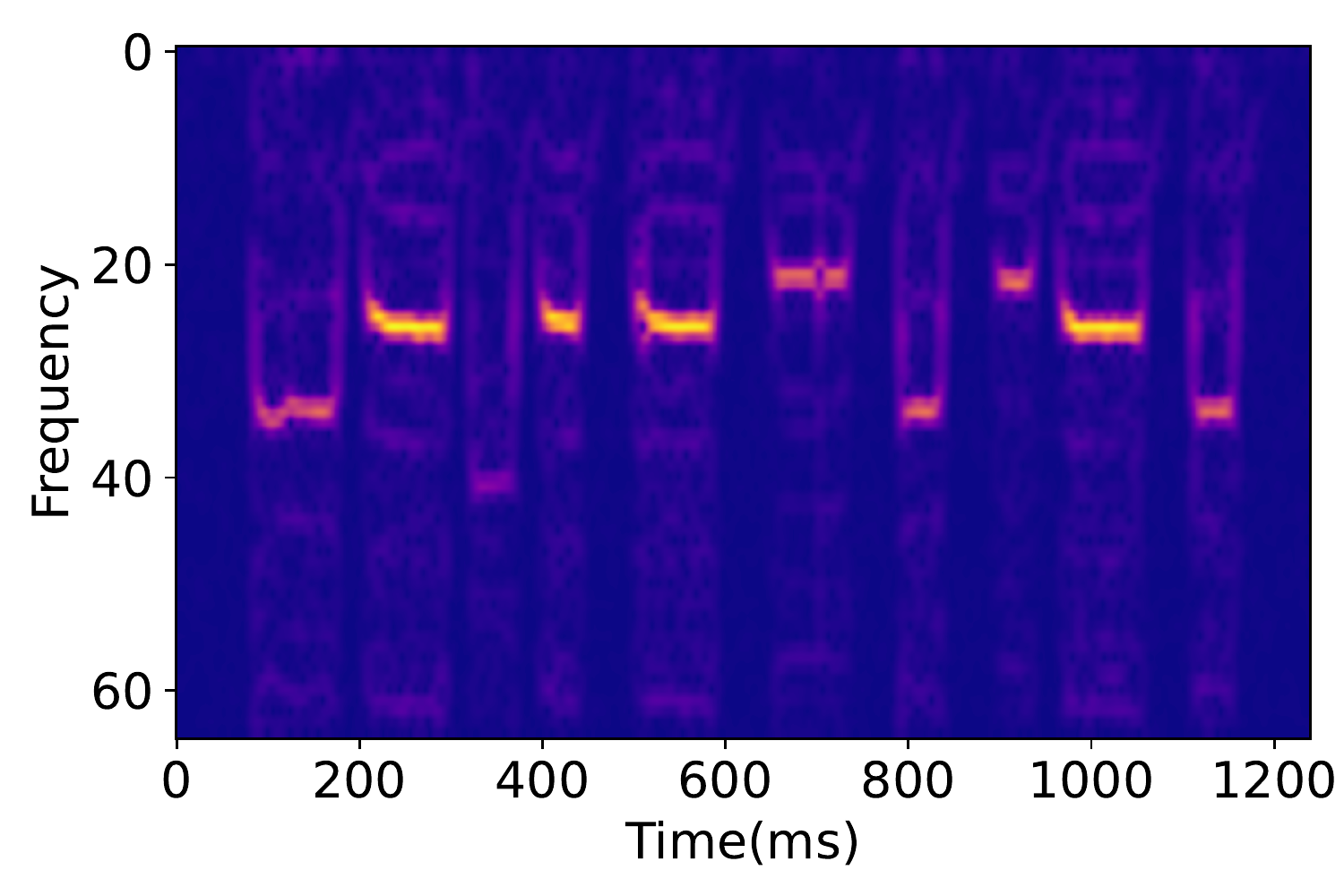}
 \caption{}
 \label{fig:spec_before}
\end{subfigure}
\begin{subfigure}[b]{0.32\linewidth}
 \centering
 \includegraphics[width=\textwidth]{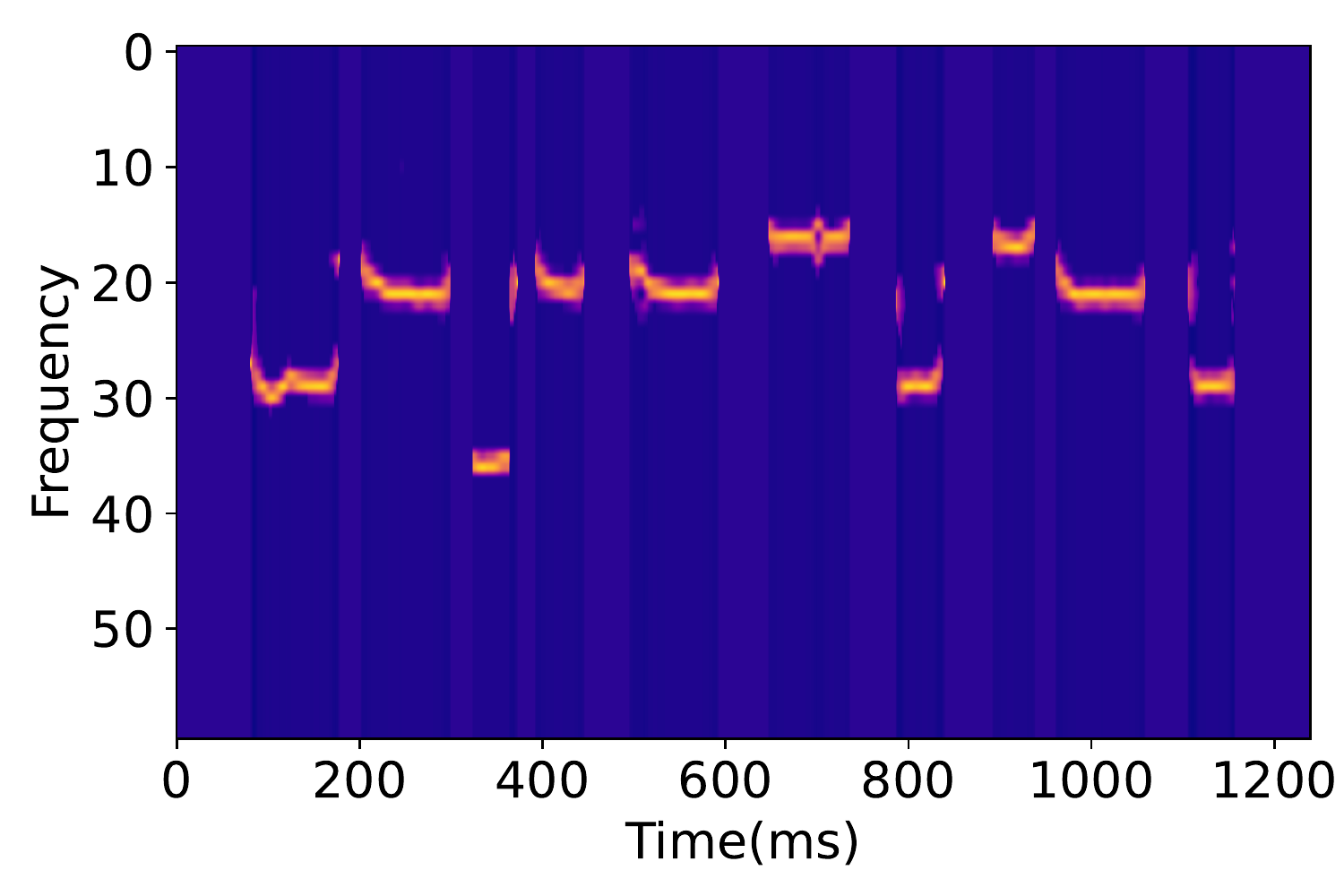}
 \caption{}
 \label{fig:spec_after}
\end{subfigure}
 \caption{(a) Raw accelerometer signal, spectrogram of the accelerometer signal (b) before filtering, (c) after filtering to remove noise.}
\label{fig:spec_before_after}
\end{figure}

\textbf{Peak Detection.}
In order to determine the \texttt{PWM} values used during encoding, we rely on the fact that the frequency of the vibrations are linked to the PWM. By accurately determining these frequencies, we can estimate the \texttt{PWM} value. 
It was observed that for most \texttt{PWM} values, there are two frequencies where there is a significant amount of energy present. This can be observed in the \cref{fig:spectrogramPWMs}, the two darkest yellow bands for each \texttt{PWM} represent these two frequencies.
They represent the two most prominent \emph{overtones} of the vibrations passing through the hand (overtones are any frequencies higher than the lowest frequency present in the signal \cite{bloothooft1992acoustics}). 
Due to the presence of two prominent overtones, in order to derive the frequency of the vibration at any point in time, our algorithm detects the two most prominent peaks in the spectrogram and computes the mean of the frequency of these two overtones to obtain a close estimation of the transmitted frequency.
However, at higher \texttt{PWMs} (e.g., 100), it was noticed that only one overtone contains a majority of the energy. In order to handle these cases, the algorithm compares the frequencies of two most prominent peaks, and only if the second most prominent peak is of the same order of magnitude as the most prominent one, the mean of the two frequencies is computed. Otherwise, it simply considers the frequency of the most prominent peak for the frequency estimation.

{\small
\noindent
\begin{algorithm}[tbph]
\caption{Symbol separation algorithm.}\label{alg_symb_sep}
\begin{algorithmic} 
\STATE \textbf{Input:} $y[]$
\STATE \textbf{Output:} $ONTimes[], OFFTimes[], PWMS[]$
\STATE \textbf{Initialization:} $ONTimes \gets [], OFFTimes \gets [], PWMS \gets [], startTimes \gets [], endTimes \gets [], i \gets 0$
\FOR {$i <= $length($y$)}
    \IF {$(y[i] >=0$ \text{and} $y[i+1],y[i+2],y[i+3]=0)$}
        \STATE $endTimes.append(i)$
    \ENDIF
    \IF {$(y[i] >=0$ \text{and} $y[i-1],y[i-2],y[i-3]=0)$}
        \STATE $startTimes.append(i)$
    \ENDIF
\ENDFOR
\STATE $ONTimes[] \gets endTimes[] - startTimes[] $
\STATE $j \gets 0$
\FOR {$j <= $length($ONTimes$)}
    \IF {$(ONTimes[j] < 200)$}
        \STATE $ONTimes.remove(j)$
    \ENDIF
\ENDFOR
\STATE $OFFTimes[] \gets startTimes[1:(length(startTimes)-1)] - endTimes[0:(length(endTimes)-1)]$
\STATE $k \gets 0$
\FOR {$j <= $length(ONTimes)}
    \STATE $PWMs[j] = mean(y[startTimes[j]:endTimes[j]])$
\ENDFOR
\end{algorithmic}
\end{algorithm}
}

\textbf{Symbol Separation.}
From the previous peak detection process, we obtain a vector (denoted as $y$) of the detected frequency for each time window. In time windows without any vibration, the frequency is set to $0$. We then use the following heuristic to separate out the transmitted symbols, in the form of \texttt{ON} times, \texttt{OFF} times and \texttt{PWMs}. First, the first and last time windows of each vibration are computed by using the \emph{symbol separation algorithm} (\cref{alg_symb_sep}). Then, the \texttt{ON} times, \texttt{PWMs} and \texttt{OFF} times are computed by using this information. In order to negate any effects of short-duration impulsive noise, we discard any symbols whose \texttt{ON} times are smaller than 200 \emph{ms}. We do this filtering because the shortest possible length of a transmitted symbol is 250 \emph{ms} as per the \texttt{ON} time values we have chosen for transmission, which implies that \texttt{ON} times smaller than 200 \emph{ms} are noise.

\textbf{Pilot Sequence based Mapping and Decoding.}
As mentioned earlier, a pilot sequence is appended to each message to improve accuracy at the time of detection. Due to a variety of reasons, such as the frequency response of the user's hand, the precise orientation, tightness of the watch, and the time taken by the motor to ramp up and ramp down, there could be considerable variance in the received parameters with every use of the communication system. 
Hence, we use this pilot signal during decoding to measure any offsets in the transmitted parameters and adjust them accordingly. 
The pilot signals consists of the following sequence of \texttt{ON} times, \texttt{OFF} times and \texttt{PWMs}, respectively: \texttt{ON} time values: [250,500,250], \texttt{OFF} times values: [150,300], \texttt{PWM} values: [20,30,60].

These encompass most of the parameters used to transmit information in our system. At the receiver, this pilot sequence first undergoes the spectrogram based filtering, peak detection based frequency estimation followed by symbol separation to identify the parameters used in the pilot sequence. 
These values obtained are then used as a mapping to decode the actual message. We start with the pilot sequence, where each value \(p_i^{pilot}\) in the estimated parameter set \(PWM^{e}\) of the pilot sequence $e$ is mapped to its corresponding value \(q_i\) from the original parameter set \(PWM^{o}=\{20,30,60\}\). By using this information, the output of the symbol separation algorithm is then mapped to their corresponding parameters as follows: For each estimated parameter \(p_i^{msg}\) in the estimated parameter set \(PWM^{msg}\) for the transmitted message $msg$, if \(p_i^{msg} < mean(p_j^{pilot},p_{j+1}^{pilot})\), \(p_i^{msg}\) is mapped to \(p_j^{pilot}\), else, \(p_i^{msg}\) is mapped to \(p_{j+1}^{pilot}\). For example, for \texttt{PWM} values of the pilot sequence, the symbol separation algorithm may output the following values, 24, 35, 66. Now since we know that pilot sequence can only have \texttt{PWM} values 20, 30 and 60 and the order of their occurrence, we map these estimated values to each corresponding \texttt{PWM} value. Then, for the actual message, if an estimated \texttt{PWM} value is 27, it is mapped to the pilot sequence estimated value of 24 (via the above described method) which corresponds to \texttt{PWM} 20 of the original encoding parameter value.
The same procedure is followed for other parameter sets, \(ON^{msg}\) and \(OFF^{msg}\). Finally, the message is decoded by reconstructing 4-bit long words for each \texttt{PWM}, \texttt{ON} and \texttt{OFF} combination of values obtained from the mapped symbol separation output, i.e., the first 2 bits of a word are derived based on the \texttt{PWM} value, while the third and fourth bits are derived using the \texttt{ON} time and \texttt{OFF} time value, respectively.

\textbf{A sample PIN transmission.}
\cref{fig:samplePIN} instantiates the aforementioned algorithms of \name with a real example, wherein a 4-digit PIN is first encoded into a vibration pattern (consisting of \texttt{ON}, \texttt{OFF}, and \texttt{PWM} parameters) and then transmitted. A spectrogram composed from the received accelerometer signal is then used to detect the relevant \texttt{ON}, \texttt{OFF}, and \texttt{PWM} parameters based on the vibration pulses, which are then used to successfully decode and output the original PIN.

\begin{figure}[h]
    \centering
    \includegraphics[width=\linewidth]{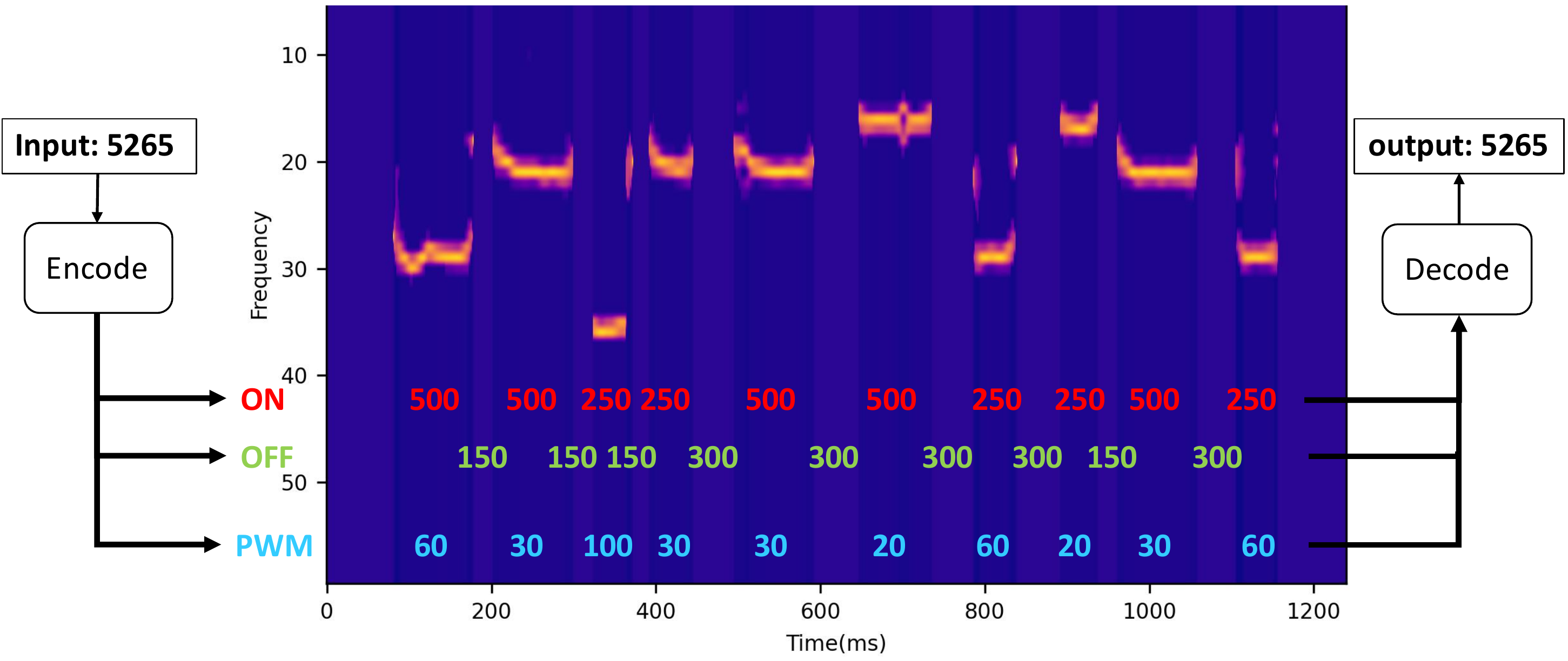}
    \caption{Encoding and decoding of a PIN.}
    \label{fig:samplePIN}
\end{figure}

\subsection{Implementation Hardware} %
\label{sub:imp_hardware}

To comprehensively evaluate \name, we implement two different hardware setups. To enable an exhaustive investigation of the performance for a variety of fine-grained transmitter and receiver parameters, we first implement and evaluate \name in a custom hardware setup (\cref{subsub:custom_hardware}). 
Then to evaluate \name in a realistic setting, we implement a consumer hardware setup (\cref{subsub:consumer_hardware}) comprising of commercially available mobile and wrist wearable devices.

\subsubsection{Custom Hardware Setup}
\label{subsub:custom_hardware}
We implement this setup by building custom devices closely resembling transmitters and receivers on commercial handheld devices (e.g., smartphones) and wrist wearables (e.g., smartwatches).
In this setup, we use Arduino Uno and Nano boards to control the transmitter and the receiver, respectively. For the transmitting vibration motor, we use an ERM motor (16000 \emph{RPM}) connected to the Arduino Uno board via a L298 motor driver \cite{L298N}, powered using a 12\emph{V} (5 \emph{Amp-Hours} capacity) Lead Acid battery. 
We use a 12V battery to power the motor to achieve longer operating times for our experiments. However, as the motor operates in the range of 1.5V to 3V it can also easily be powered using two (1.5V) AAA batteries.
We power the Arduino Nano board (the receiver device) via a 5\emph{V} USB input. To build the custom handheld device prototype, we use a consumer level smartphone case (\cref{fig:watchOrientation}) on which we mount the vibration motor. For the custom wrist wearable prototype, we use a Sony Smartwatch 3 Band (the smartwatch portion removed, see \cref{fig:watchOrientation}) to mount the motion sensor.
For both the custom devices, we use a MPU6050 GY-521 MEMS motion sensor containing a three-axis accelerometer and a three-axis gyroscope.
The motion sensor is also connected with a Micro-SD card using an Arduino Micro-SD card adapter (via SPI mode) for data recording purposes.
As explained earlier (\cref{sec:background}), we use a \texttt{PWM} technique to control the operation of vibration motor. The sampling rate of the motion sensor achieved when writing to the Micro-SD card is 700 \emph{Hz}. To evaluate communication in the reverse direction (i.e. wrist wearable to handheld device), we swap the motor and motion sensor between the two custom devices.

\begin{figure}[h]
\begin{subfigure}[b]{0.59\linewidth}
 \includegraphics[width=\textwidth]{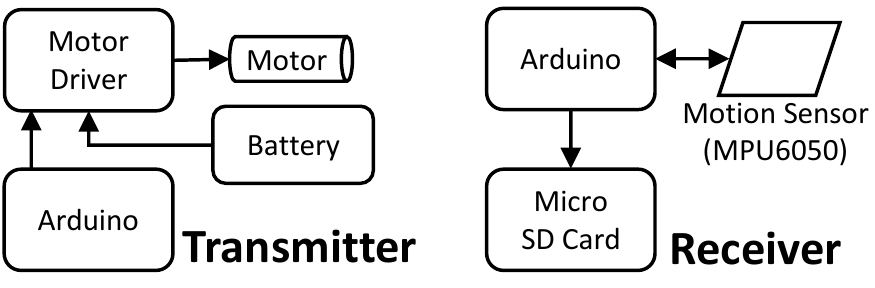}
 \caption{}
 \label{fig:transmitterReceiverArchi}
\end{subfigure}
\hfill
\begin{subfigure}[b]{0.3\linewidth}
 \includegraphics[width=\textwidth]{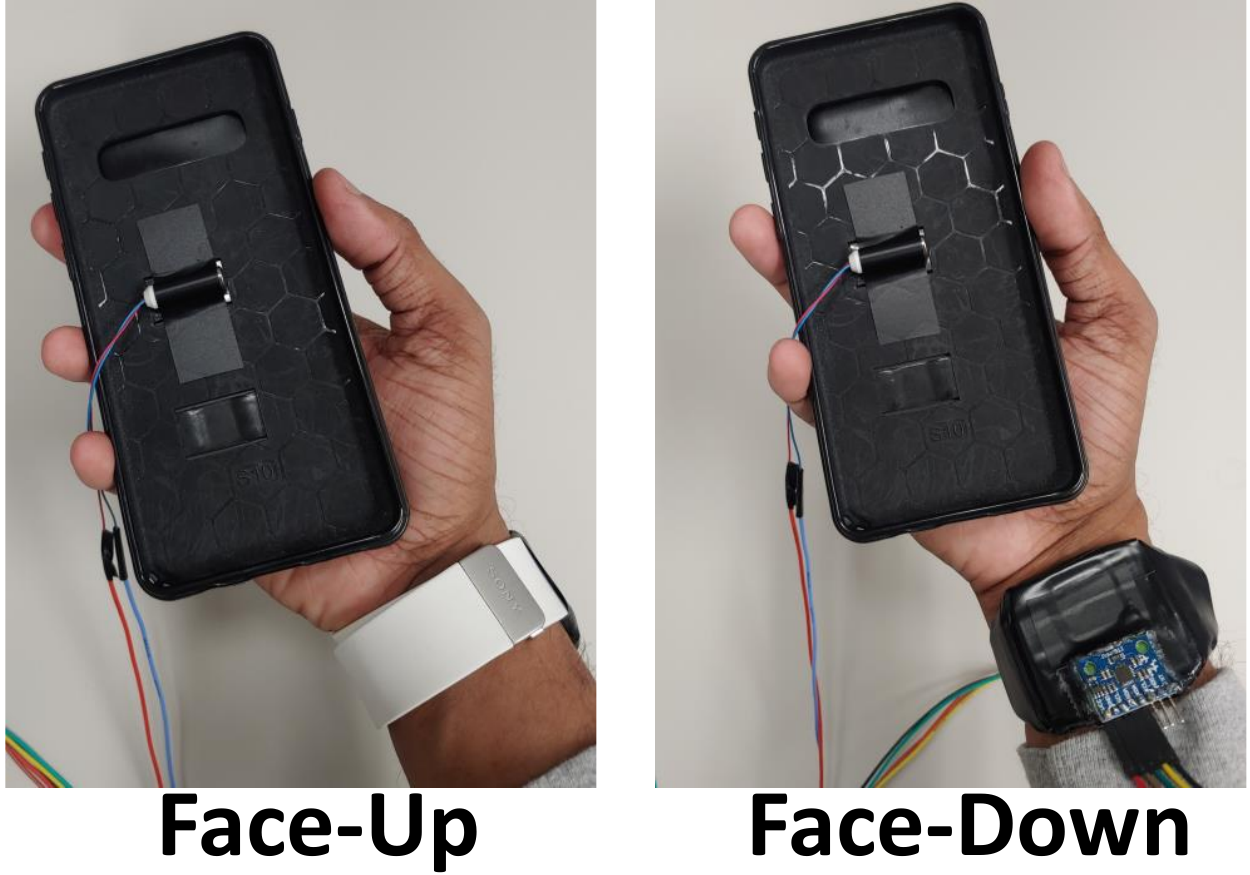}
 \caption{}
 \label{fig:watchOrientation}
\end{subfigure}
\caption{(a) Custom hardware architecture, (b) wrist wearable wearing orientations.}
\label{fig:circuit_diagram}
\end{figure}

\subsubsection{Consumer Hardware Setup}
\label{subsub:consumer_hardware}
For this setup, we use a Nokia 6 (2017) smartphone as the handheld device and a Sony Smartwatch 3 as the wrist wearable. The Sony Smartwatch can only achieve a maximum sampling rate of 200 \emph{Hz}, in contrast to the sampling rate of 700 \emph{Hz} that we were able to achieve using the custom hardware setup. Moreover, in these devices the Android API only provides control of the motor's vibration amplitude, but not of the frequency. Similar to Roy et al. \cite{roy2015ripple}, we looked into the possibility of modulating the amplitude to encode data. However as mentioned earlier, due to sampling rate limitations in consumer level smartwatch accelerometers, we were not able to accurately differentiate between different amplitudes just by using the accelerometer data. This effectively limits the full implementation of the \name communication protocol on these devices as we rely on multiple frequencies to encode data.

\subsection{Human Subject Data Collection} %
\label{sub:data_collection}
We perform a comprehensive empirical evaluation of \name using data collected from human subject participants in realisitic settings and environments.
For our data collection study, we recruited 13 participants in the age group of 18-29 from our university campus. Participants in our study wore the custom wrist wearable device on the wrist of one hand (preferred by the participant) and held the custom handheld device (\cref{fig:watchOrientation}) in the palm of the same hand. In this setup, binary data was communicated from the handheld device to the wrist wearable, and vice versa, using our proposed vibration-based approach under a variety of different ambient/device settings and algorithm parameters.

In our first experimental setup, we evaluate the effect of two different orientations of the wrist wearable device on the performance of the communication scheme. The first orientation is the case when the wrist wearable is worn facing upwards along the top of the forearm (or wrist), while the second orientation is when the wrist wearable is facing downwards along the bottom of the wrist (see \cref{fig:watchOrientation}). 
As different users may prefer to wear their wrist wearables in different orientations and, depending on the orientation of the wrist wearable the position of the vibration motor and accelerometer on the wrist/forearm may change impacting the performance of \name, it is important to further analyze these setups. 
In the second experimental setup, we test the effect of distance between the transmitter (vibration motor) and receiver (accelerometer) devices on \name's performance. Accordingly, we mount the motor at three different positions on the custom handheld device, top (15 \emph{cm}), middle (7.5 \emph{cm}) and bottom (2 \emph{cm}). In the third data collection setting, participants were asked to walk while holding the custom handheld device and wearing the wrist wearable while the devices were communicating with each other. The purpose of this setting was to study how noise introduced (in the accelerometer sensor readings) due to physical activities such as walking impacts the performance of \name. In the fourth experimental setup, we tested the reverse direction, in which the wrist wearable (mounted with a motor) acts as the transmitter and the handheld device (mounted with an accelerometer) act as the receiver. This setup would evaluate the half-duplex property of \name.
Finally, the different orientation and participant walking setups are repeated for the reverse communication direction as the fifth and sixth experimental setup.
In addition to these experiments, we also conducted an experiment to test \name using consumer level devices where a set of participants wore a Sony Smartwatch 3 on their wrist while holding a Nokia 6 (2017) smartphone on the palm of the same hand. Similar to the above experiments, binary data was communicated from the handheld device to the wrist wearable. 
In each of these experiments, four 128-bit randomly generated binary strings are used as test communication (transmission). All data collection was done after obtaining consent from the participants, and our data collection and analysis procedures were approved by our university's Institutional Review Board (IRB).

%% file: evaluation.tex
\section{Evaluation} %
\label{sec:evaluation}
In this section, we present a comprehensive empirical evaluation of \name under a variety of operational settings, algorithm parameters and hardware setups.
We evaluate the accuracy and efficiency of \name using the following standard metrics: (i) Bit Error Rate (\emph{BER}): \emph{BER} is the ratio of the number of incorrectly interpreted bits, (ii) Bit Rate (\emph{BR}): \emph{BR} is the transmission speed, i.e. the number of bits transmitted per unit time (seconds).

\begin{figure*}[] 
     \centering
     \begin{subfigure}[b]{0.24\linewidth}
         \centering
         \includegraphics[width=\textwidth]{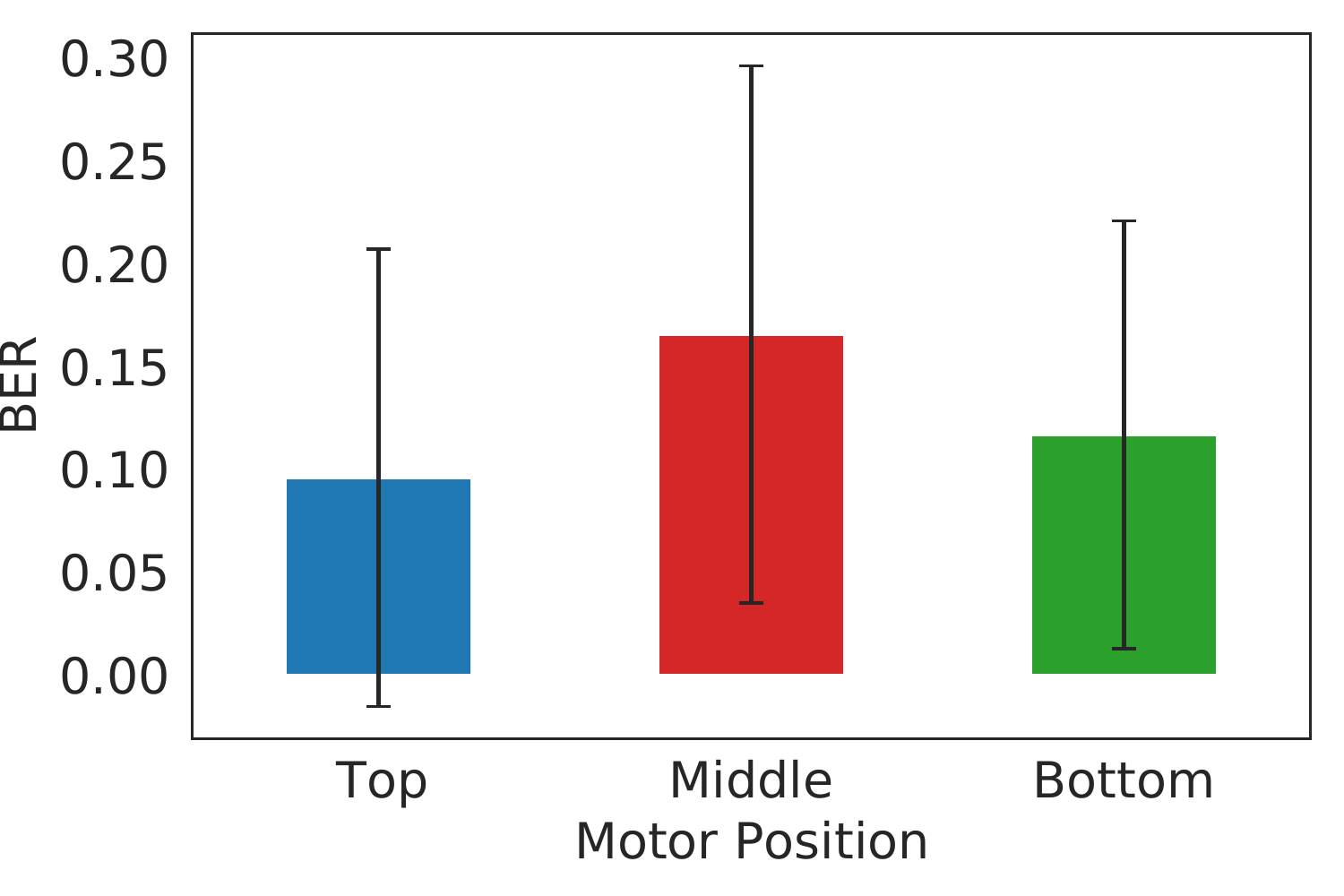}
         \caption{}
         \label{motor_position}
     \end{subfigure}
     \hfill%
     \begin{subfigure}[b]{0.24\linewidth}
         \centering
         \includegraphics[width=\textwidth]{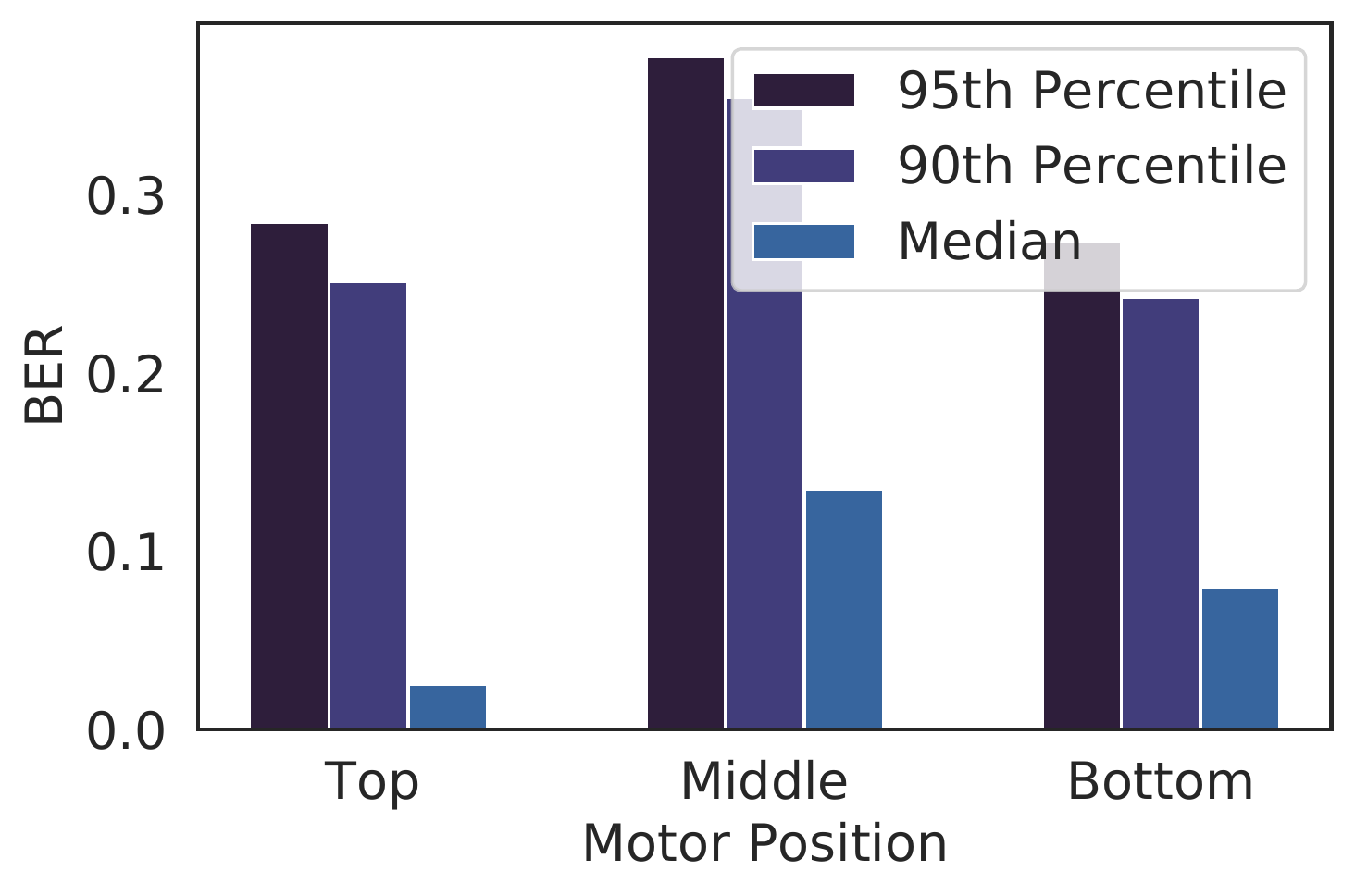}
         \caption{}
         \label{distance_percentile}
     \end{subfigure}
     \hfill%
     \begin{subfigure}[b]{0.24\linewidth}
         \centering
         \includegraphics[width=\textwidth]{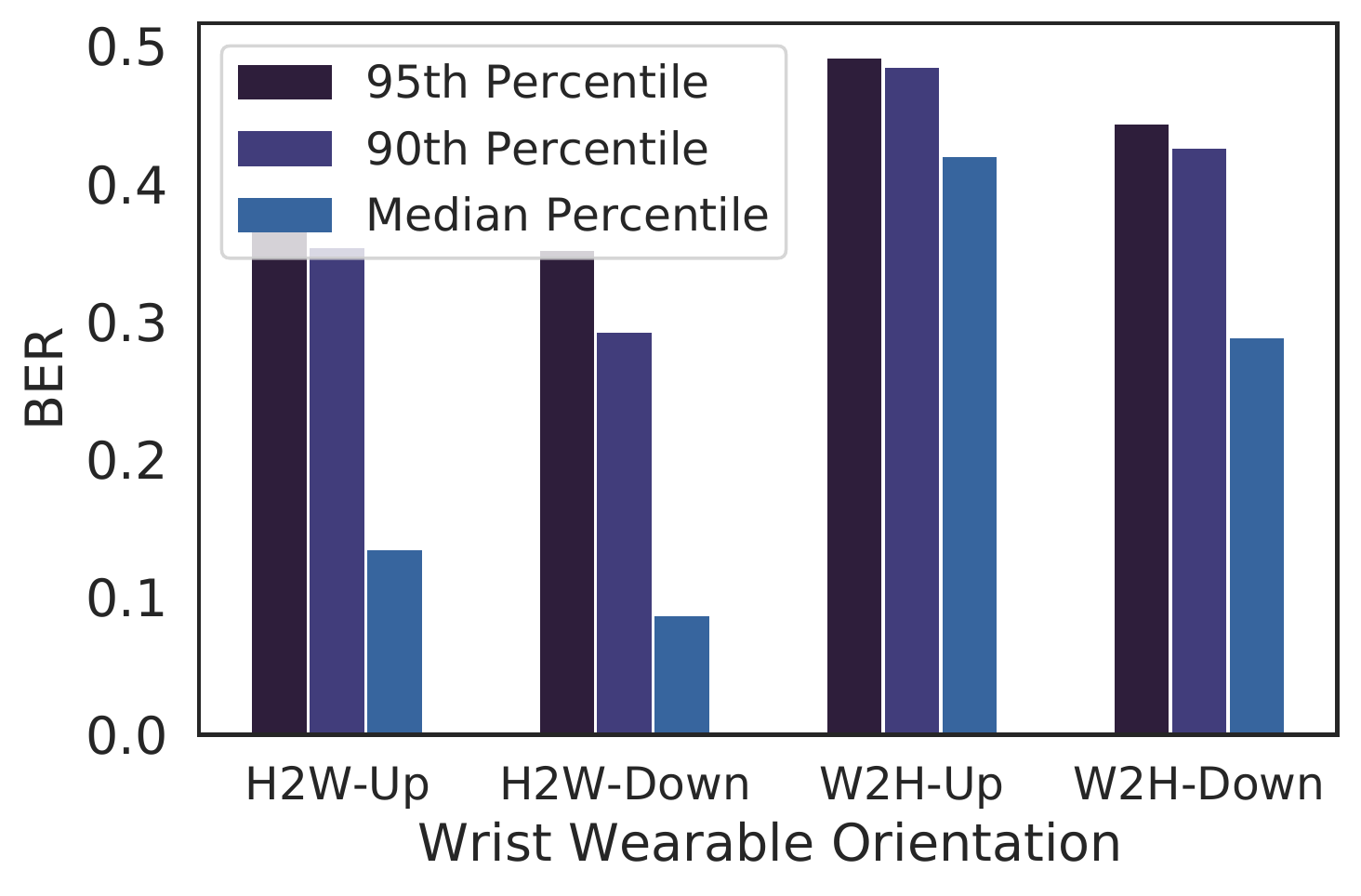}
         \caption{}
         \label{watch_face_percentile}
     \end{subfigure}
      \hfill%
     \begin{subfigure}[b]{0.24\linewidth}
         \centering
         \includegraphics[width=\textwidth]{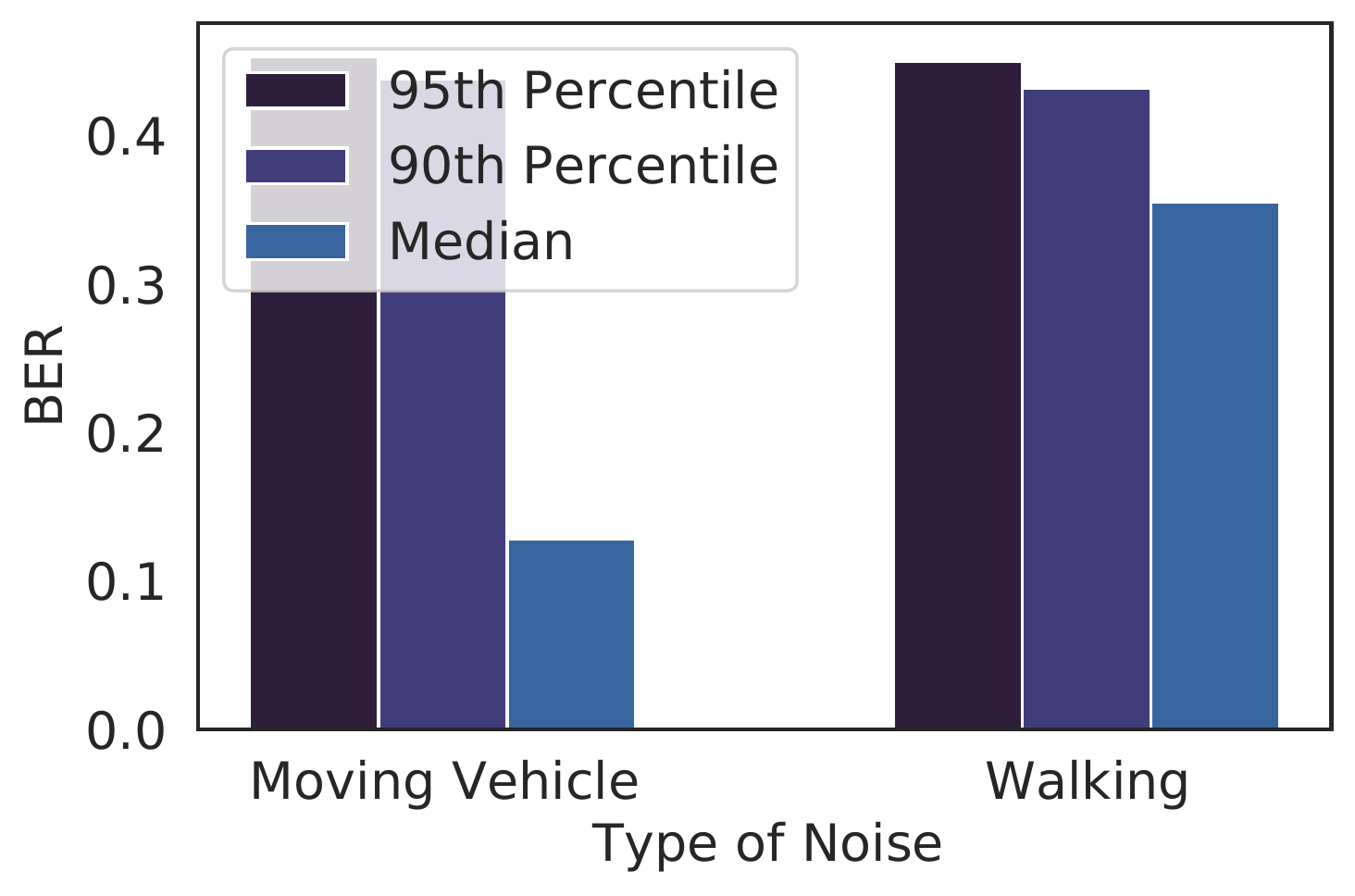}
         \caption{}
         \label{noice_percentile}
     \end{subfigure}
     \begin{subfigure}[b]{0.24\linewidth}
         \centering
         \includegraphics[width=\textwidth]{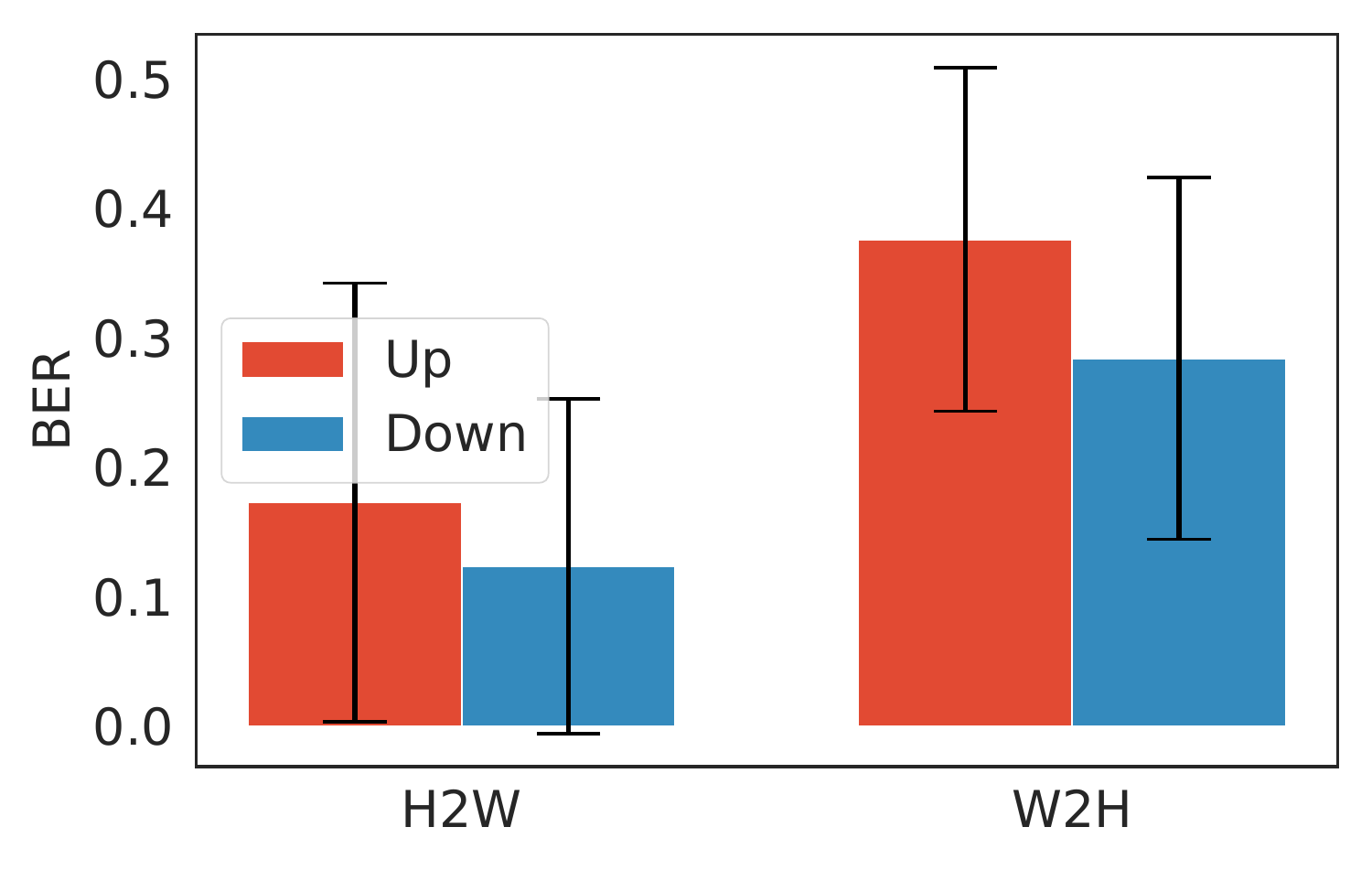}
         \caption{}
         \label{watch_face}
     \end{subfigure}
      \hfill%
     \begin{subfigure}[b]{0.24\linewidth}
         \centering
         \includegraphics[width=\textwidth]{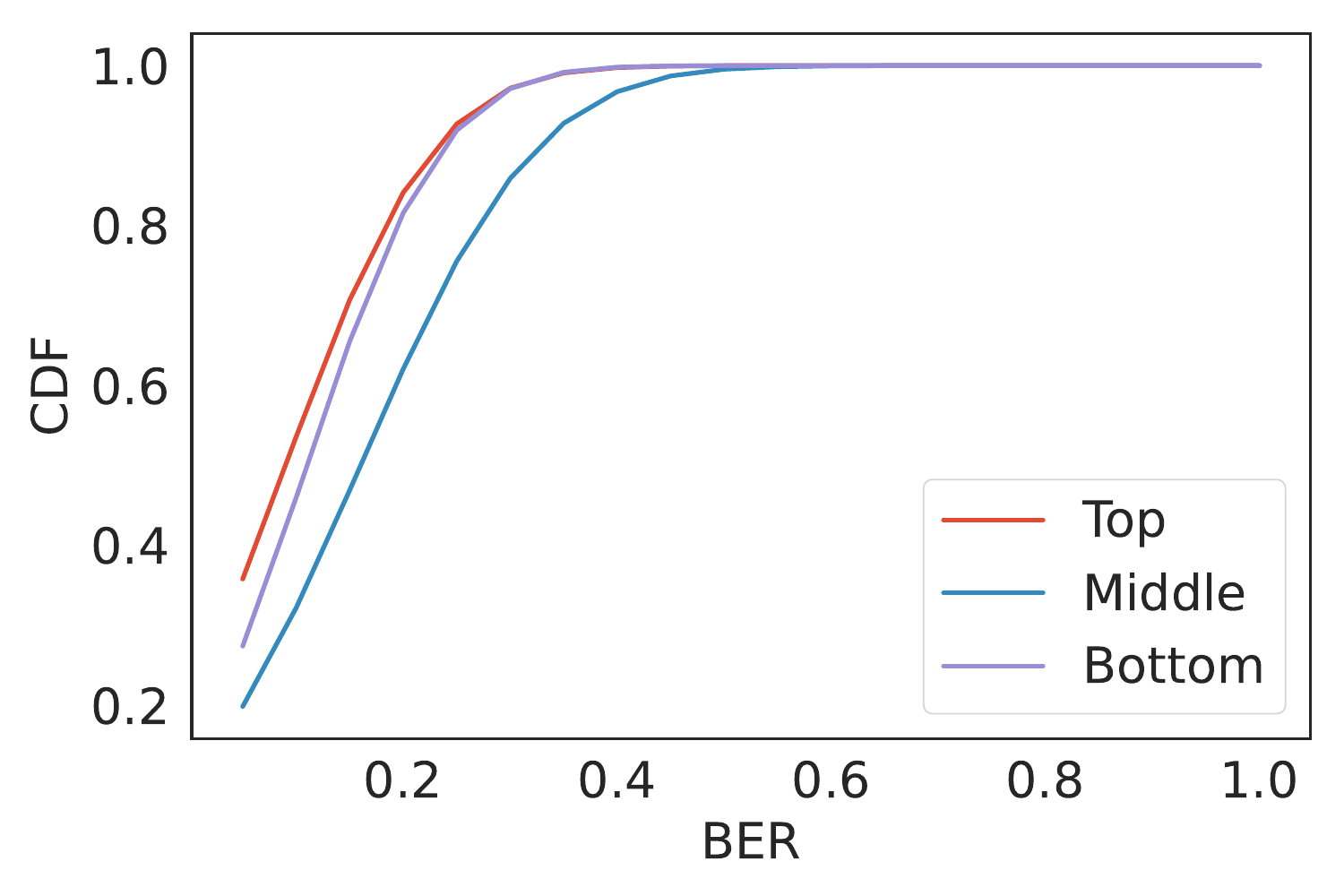}
         \caption{}
         \label{distance_cdf}
     \end{subfigure}
      \hfill%
     \begin{subfigure}[b]{0.24\linewidth}
         \centering
         \includegraphics[width=\textwidth]{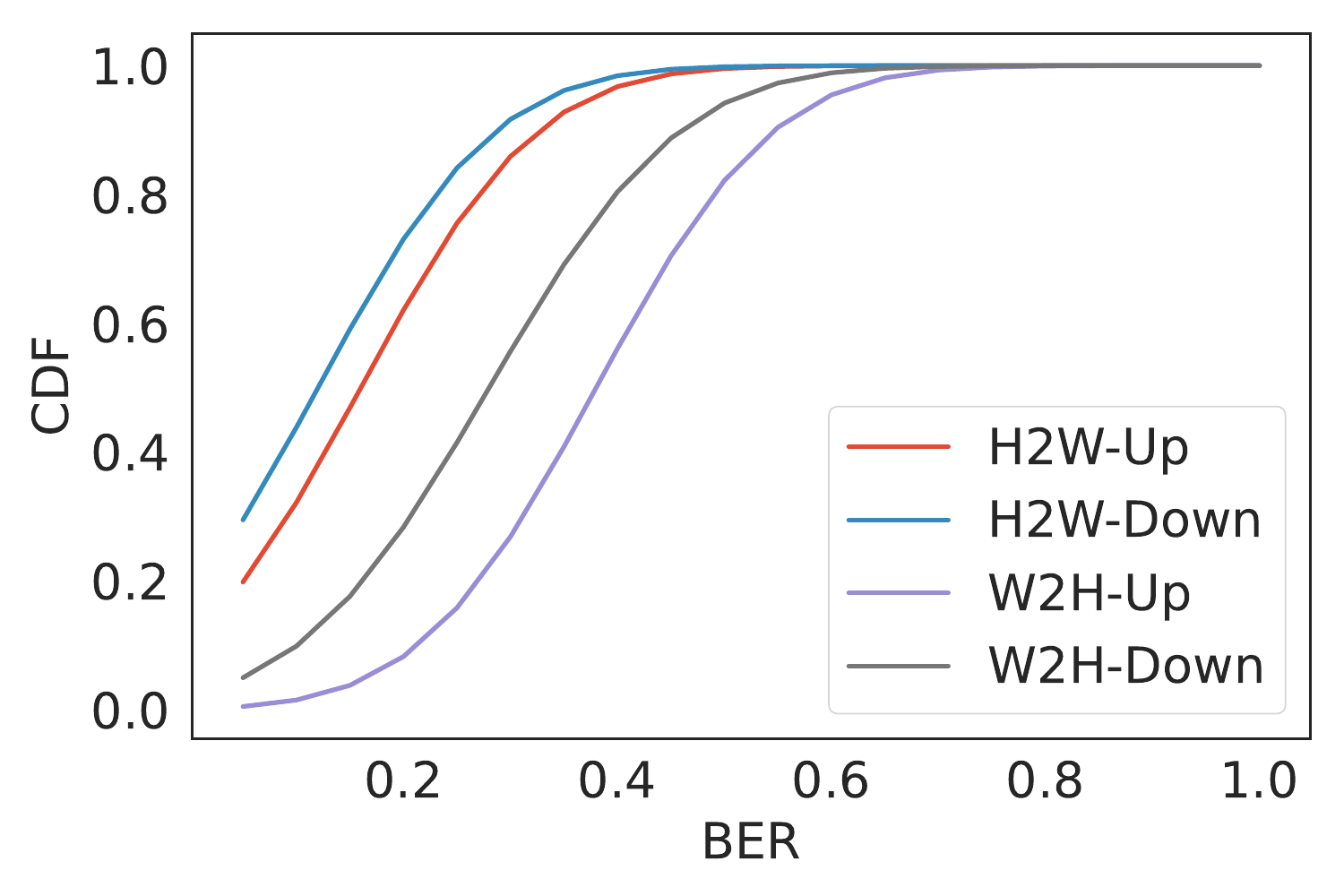}
         \caption{}
         \label{orientation_cdf}
     \end{subfigure}
      \hfill%
     \begin{subfigure}[b]{0.24\linewidth}
         \centering
         \includegraphics[width=\textwidth]{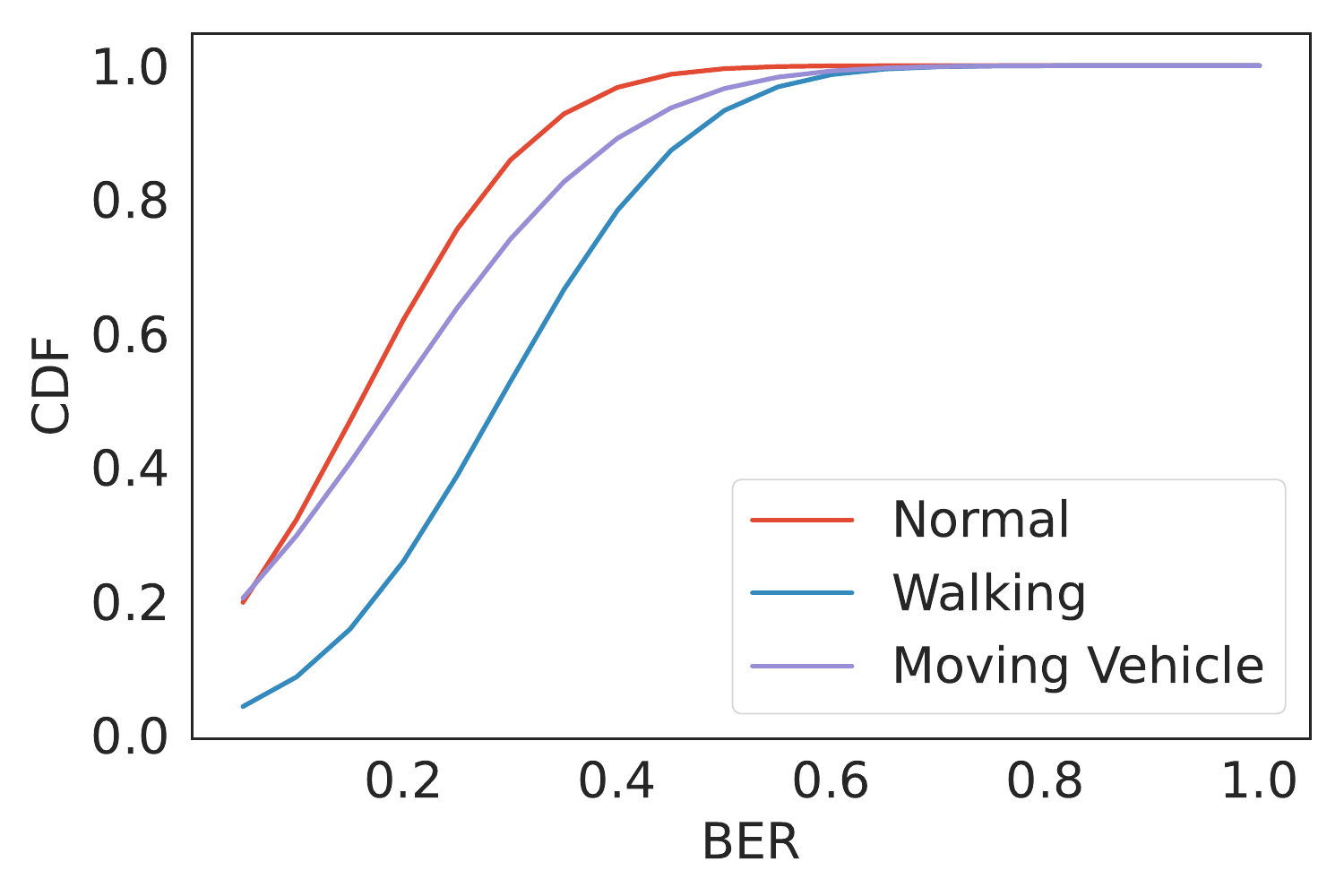}
         \caption{}
         \label{noise_cdf}
     \end{subfigure}
         \caption{Mean BERs: (a) distance between the motor and the motion sensor, (e) wrist wearable orientation. BERs of participants at percentiles $95^{th}$, $90^{th}$, and median: (b) distance between the motor and the motion sensor, (c) wrist wearable orientation, (d) effect of different types of noise. CDF of BER: (f) distance between the motor and the motion sensor, (g) wrist wearable orientation, (h) effect of different types of noise. }
        \label{fig:motor_position_plots}
\end{figure*}

\subsection{Effect of Distance Between Transmitter (Motor) and Receiver (Motion Sensor)} %
\label{sub:distance_between_watch_and_phone}

In order to observe the effect of distance between the transmitter (vibration motor) and receiver (accelerometer sensor) on system performance, we mount the motor on the phone case at three different positions. Specifically, the motor is mounted at the top, middle and bottom positions of the handheld device case, which results in a distance of approximately 15 \emph{cm}, 7.5 \emph{cm}, and 2 \emph{cm} to the wrist wearable, respectively. We observe that when the motor is at the middle position, the performance slightly degrades (see \cref{motor_position}), while for top and bottom positions the resulting mean \emph{BERs} are comparable.
We further observe that in the middle motor position, for half the participants the resulting \emph(BER) values are lower than 0.15 (see \cref{distance_percentile}) with five of them having \emph{BERs} lower than 0.05. For the top (farthest from the receiver) motor position, we observe the best mean \emph{BER} of 0.10 ($\sigma=0.13$) and that half the number of participants have \emph{BERs} lower than 0.05. In the bottom position, we have similar performance with a mean \emph{BER} of 0.12 ($\sigma=0.10$) and 50\% of the participants with a \emph{BER} less than 0.10.
We observed that when the motor is positioned at the top of the handheld device, although farthest from the receiver (accelerometer), the vibrations makes the whole handheld device slightly more agitated.
This, in turn, results in a stronger vibration signal reaching the accelerometer at the receiver. A similar effect is observed when the motor is mounted on the bottom of the handheld device. When the motor is mounted on the middle of the device, the vibrations are slightly dampened, accompanied by the fact that human palm is curved in the middle and thus the direct surface area that may be coming in contact with the hand when vibrating is smaller, which is reflected in the slightly higher \emph{BER} values. From \cref{distance_cdf}, which depicts the cumulative probability distribution of \emph{BER} for each transmitter-receiver distance, we can see that for nearly 70\% of the cases in the top and bottom positions, the achievable \emph{BER} is less than 0.15.

\subsection{Effect of Wrist Wearable Orientation} %
\label{sub:watch_orientation}
Next we evaluate the effect of the wrist wearable's orientation on system performance, with the communication direction from the handheld device to the wrist wearable (\cref{fig:watchOrientation}). Here, we see (\cref{watch_face}) slightly lower \emph{BERs} when the wrist wearable is worn facing downwards (accelerometer located at the bottom of the wrist) compared to when facing upwards (accelerometer located on the top of the wrist), with \emph{BER} values of 0.12 ($\sigma=0.13$) and 0.16 ($\sigma=0.13$), respectively. \cref{watch_face_percentile} shows that half the participants had \emph{BERs} less than 0.10 for the face down orientation. 
We believe that this is because when the vibration motor and motion sensors are aligned on the same side of the hand, the vibrations have a more direct path to travel.
Further, from \cref{orientation_cdf}, we see that for the face down orientation, close to 70\% of the time \emph{BER} will be lower than 0.2, as opposed to 60\% probability in the face up orientation.

\subsection{Performance Under Noise}%
\label{sub:performance_under_noise}

Next we evaluate \name under two different types of movement noise that impact motion sensor readings.
The first type of noise occurs when the user is relatively stationary while traveling inside a moving vehicle (with vehicle speeds between 16 $kmh$ to 64 $kmh$).
Due to practical limitations, we simulated this experiment by superimposing prerecorded motion sensor data recorded from the handheld device while inside a moving vehicle, to the motion data collected from participants in a lab setting. 
The second type of noise we experimented with occurs due to user movement, where participants were asked to walk with the handheld device and wrist wearable on the same hand while \name is executing.
The first noise scenario (moving vehicle) shows slightly better performance with a mean \emph{BER} of 0.19 ($\sigma=0.17$). Further, 7 out of the 13 participants showed \emph{BERs} less than 0.10 (\cref{watch_face_percentile}), with only 3 participants showing significant degradation of \emph{BERs} of over 0.3. This indicates that \name is fairly robust against movement noise resulting from \textit{plane motion}, such as traveling inside a vehicle.
For the second type of noise scenario (walking), the \emph{BER} considerably degrades with an observed mean \emph{BER} of 0.29 ($\sigma=0.28$). For both the above scenarios, the communication direction was from the handheld to the wrist wearable.
Moreover, with built-in APIs \cite{androidSensor} about user movements on modern mobile operating systems, such as Android and iOS/watchOS, it is also possible to contextually turn off \name when the user is not stationary.

\subsection{Effect of Sampling Rates} %
\label{sub:sampling_rates}

To evaluate the impact of reduced motion sensor sampling rates on protocol performance, we test \name with the accelerometer (at the receiver) sampled at 200 \emph{Hz}, compared to the 700 \emph{Hz} sampling rate which was used previously. We choose a sampling rate of 200 \emph{Hz} because it is the maximum sampling rate that is achievable on most modern consumer level mobile and wearable devices. 
Based on the results obtained in this setting, we observe that the \emph{BER} of the communication scheme degrades to around 0.4 ($\sigma=0.06$) when a lower sampling rate is used. 
This indicates that such a low accelerometer sampling rate at the receiver is probably not sufficient for capturing the range of vibration frequencies that we are utilizing to encode data in our transmission algorithm. With the limitation of not being able to use multiple frequencies, achieving a better \emph{BER} would only be possible at the cost of a reduced bit rate.

\subsection{Half-Duplex Communications}
In the experimental settings discussed so far, we have only considered communication in the direction from the handheld device to the wrist wearable.
However, as modern wearable and handheld devices come equipped with both a vibration motor and motion sensors, it would be extremely useful to evaluate communication in the reverse direction to what we have evaluated so far (i.e., from the wrist wearable to the handheld, in our case). If feasible, it would grant a half-duplex property to the communication channel, which would enable a whole set of applications that require communication in both directions. As our proposed communication protocol and hardware setup is amenable to such an evaluation, we also test communications in the reverse direction, i.e. using the wrist wearable as a transmitter and the handheld device as a receiver. Overall, the achieved performance results are significantly lower compared to the earlier case, as seen in the \cref{watch_face} and \cref{watch_face_percentile} with overall \emph{BERs} dropping to 0.38 ($\sigma=0.13$) for when wrist wearable is facing upwards and 0.28 ($\sigma=0.14$) for when the wrist wearable is facing downwards. 
Similar to the wrist wearable orientation experiment (\cref{sub:watch_orientation}), we observe that when the wrist wearable is facing downwards a lower \emph{BER} can be achieved.
We also conduct the same noise-related experiments (similar to \cref{sub:performance_under_noise}) for this setting and observe a similar pattern in the performance results where the user movement (walking) experiment degraded the \emph{BER} to 0.43 ($\sigma=0.02$) while the vehicle movement experiment reduced it to 0.42 ($\sigma=0.13$).
These low performance results can be attributed to the fact that the wrist wearable surface area in contact with the wrist/hand is much smaller compared to the handheld device. This may result in reduced perception of the generated vibrations by the the skin/hand, thus resulting in an overall reduced performance.

\subsection{Consumer-grade Hardware Setup} %
\label{sub:consumer_level_hardware_setup}
Next we discuss the performance of \name when using commercial consumer-grade hardware (as summarized in \cref{subsub:consumer_hardware}). With the highly constrained access to the vibration motor and motion sensors on commercial mobile and wearable devices, \name when executed on these devices could only achieve low overall \emph{BERs} of around 0.4, for both the watch wearing orientations.
We believe that one of the main constraints is the software-restricted sampling rates of motion sensors in commercial smartphones/smartwatches, which limits the maximum allowable sampling rate to only 200 \emph{Hz}. Further, the Android API also restricts the frequency modulation of the vibration motor on these devices, which prevents us from using the embedded motor at its full operating capacity. These factors restrict us from using the 4 PWM frequency bands (of the motor) for communication which we used in the custom hardware case. As a result, we are not able to modulate in the PWM frequency band in the commercial device case, which effectively brings down the achievable bit rate from 6.6 to 3.3 \emph{bps}. 
Although the reduced performance of \name can be primarily attributed to the software and hardware limitations of existing commercially-available mobile device hardware, we believe that better motion sensors (with high sensitivity and sampling rates) and vibration motors in future devices will result in a slightly more favorable outcome for such vibration-based communication systems.

\subsection{Accelerometer vs. Gyroscope} %
\label{sub:subsection_name}
Our preliminary experiments involving both the custom and consumer-grade devices demonstrated that accelerometers produced better feedback than gyroscopes in our setup. 
To comprehensively compare the impact on performance when a gyroscope is used as a receiver as opposed to an accelerometer, we perform some additional experiments using our custom hardware setup. Based on the observed results (see \cref{acc_vs_gy}), we can conclude that our protocol produces better performance (lower \emph{BER}) using the accelerometer. The achievable \emph{BER} drops from 0.16 to 0.39 when a gyroscope is used as a receiver (when transmitting from the handheld device). In contrast to accelerometers, gyroscopes measure a device's angular velocity and it is likely that surface vibrations, which are already dampened down as they travel through the human skin/body, do not produce a significant amount of angular motion to be discernible on a gyroscope.

\begin{table}[h]
\centering
\caption{Mean BERs for accelerometer and gyroscope.}
\label{acc_vs_gy}
\begin{adjustbox}{width=0.8\linewidth}
\begin{tabular}{lll} 
\toprule
Direction & Accelerometer & Gyroscope \\ 
\hline
Handheld to wrist wearable & 0.16 ($\sigma=0.13$) &   0.39 ($\sigma=0.09$) \\ 
Wrist wearable to handheld & 0.38 ($\sigma=0.13$) &   0.41 ($\sigma=0.10$) \\
\bottomrule
\end{tabular}
\end{adjustbox}
\end{table}

\subsection{Failed Transmission Detection} %
\label{sub:failed_transmission_detection}
After further scrutinizing the transmitted messages with higher \emph{BERs}, we observed that these inaccuracies are mostly caused due to missing bits, which may be due to voluntary or involuntary hand movements occurring during data transmission. To overcome this, we propose a message length based verification at the receiver. For this, if we assume that the receiver knows the length of the incoming message, or that the message length is fixed. Incorrectly or erroneously received messages can be easily identified (and flagged) based on bit length mismatches.
In case of such a mismatch detection, the receiver can request a re-transmission. We observe that using such a heuristic significantly improves the \emph{BERs} in our custom hardware setup, where \emph{BERs} dropped below 0.04 when erroneously received messages are correctly identified for re-transmission.

\subsection{In-Depth Analysis} %
\label{ssub:in_depth_analysis_of_the_error_rates}

\subsubsection{Bit Rate vs Error Rate} %
\label{ssub:bit_rate_vs_error_rate}

\begin{figure}[htbp]
\begin{subfigure}[b]{0.49\linewidth}
    \includegraphics[width=\linewidth]{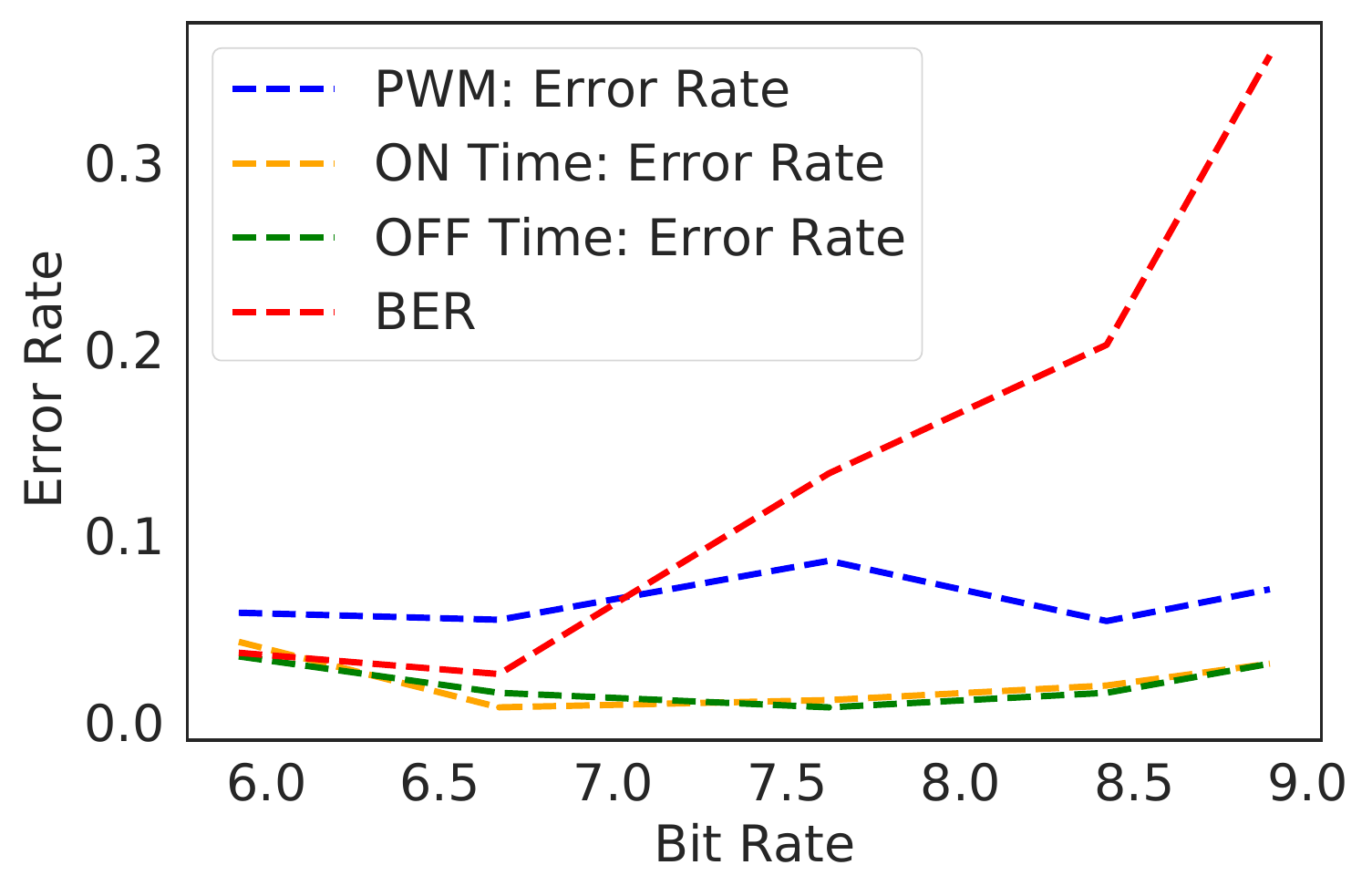}
    \caption{}
    \label{mode_er_br}
\end{subfigure}%
\hfill%
\begin{subfigure}[b]{0.49\linewidth}
    \includegraphics[width=\linewidth]{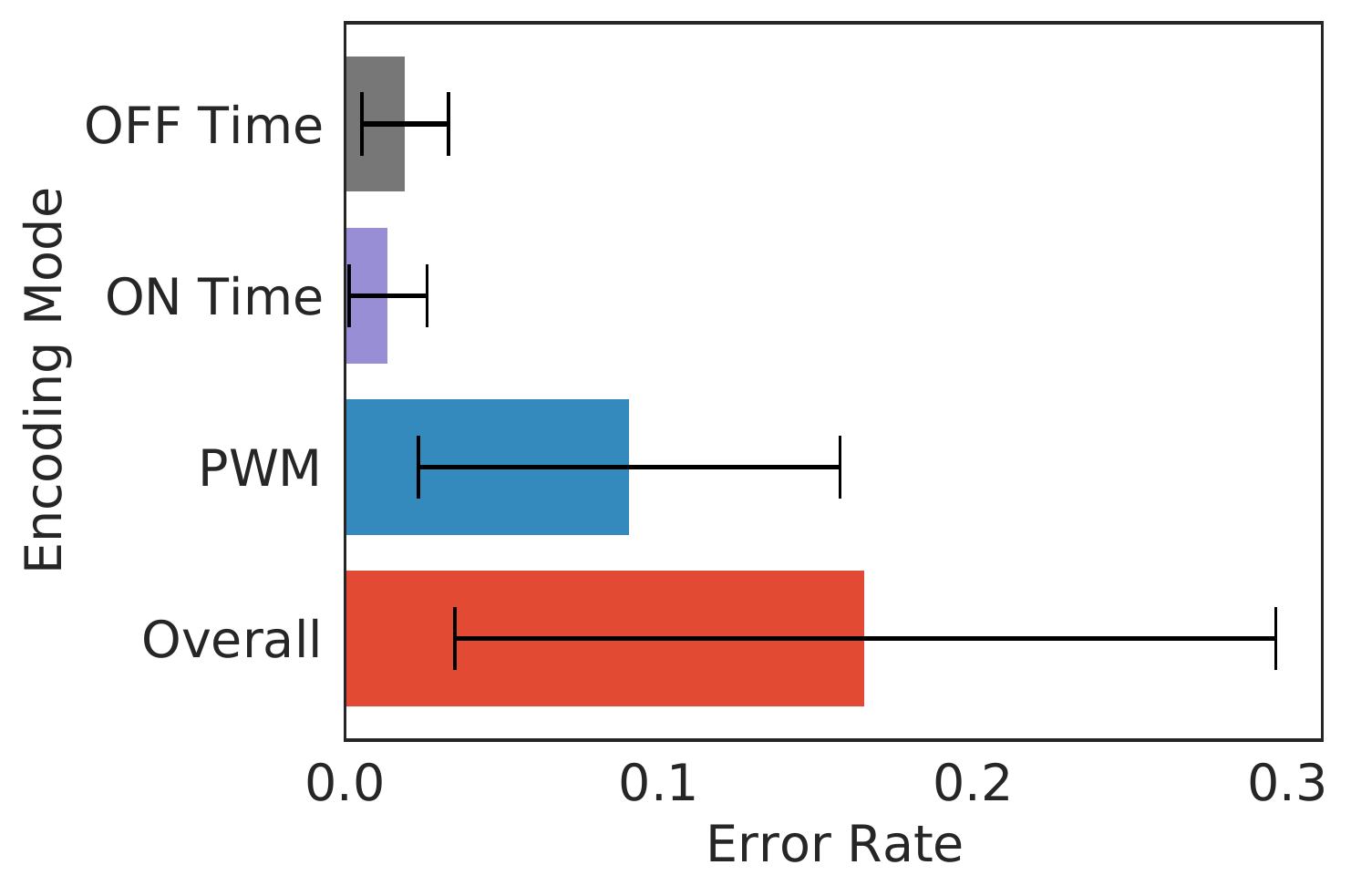}
    \caption{}
    \label{mode_overall_er}
\end{subfigure}
\caption{Individual encoding parameters (a) Error rates vs bit rate (bit/s), (b) Error rate comparison.}
\end{figure}

As discussed before, to gain higher transmission speeds we can reduce the \texttt{ON} times of the encoding algorithm, but this directly affects the accuracy which can be seen in \cref{mode_er_br}. At higher bit rates, starting around 7 \emph{bps}, \emph{BER} steadily increases. However, we believe that with the use of more advanced, highly sensitive motion sensors that could be operated at higher sampling rates, \name could achieve higher transmission speeds while maintaining a low error rate.

\cref{conf_m} shows the confusion matrix for PWM-based symbols, and we observe that most errors occur between adjacent PWMs (e.g. 20 and 30, or 60 and 100). As previously discussed in \cref{ssub:transmitter_design}, we chose the PWMs to be 20, 30, 60, and 100 after observing that they have minimal confusion with each other when decoded via the accelerometer signal. However, when testing under realistic settings with possible variations in bone structure of hands of different users, along with variations in the way they hold a handheld device in their palm, we observed that some of these PWM-based frequency vibrations could get picked up by an accelerometer differently. Essentially, if we are able to use additional PWMs (i.e. vibration frequencies) as carriers to encode data, we would be able to achieve a higher bit rate. But due to the fact that accelerometer sensors are not able to distinctly identify some of the adjacent frequencies, it limits us from achieving a higher bit rate. In other words, higher bit rates would come at the cost of higher error rates (\emph{BERs}). This is further confirmed by Roy et al. \cite{roy2015ripple}, who also highlighted in their work that high energy vibrations, occurring in the resonant frequency band, could potentially interfere with neighboring frequencies, which also limit the number of usable frequency bands in such vibration-based communication schemes.

\subsubsection{Analysis of Individual Encoding Parameters}

\begin{figure}[htbp]
\begin{subfigure}[b]{0.38\linewidth}
    \includegraphics[width=\linewidth]{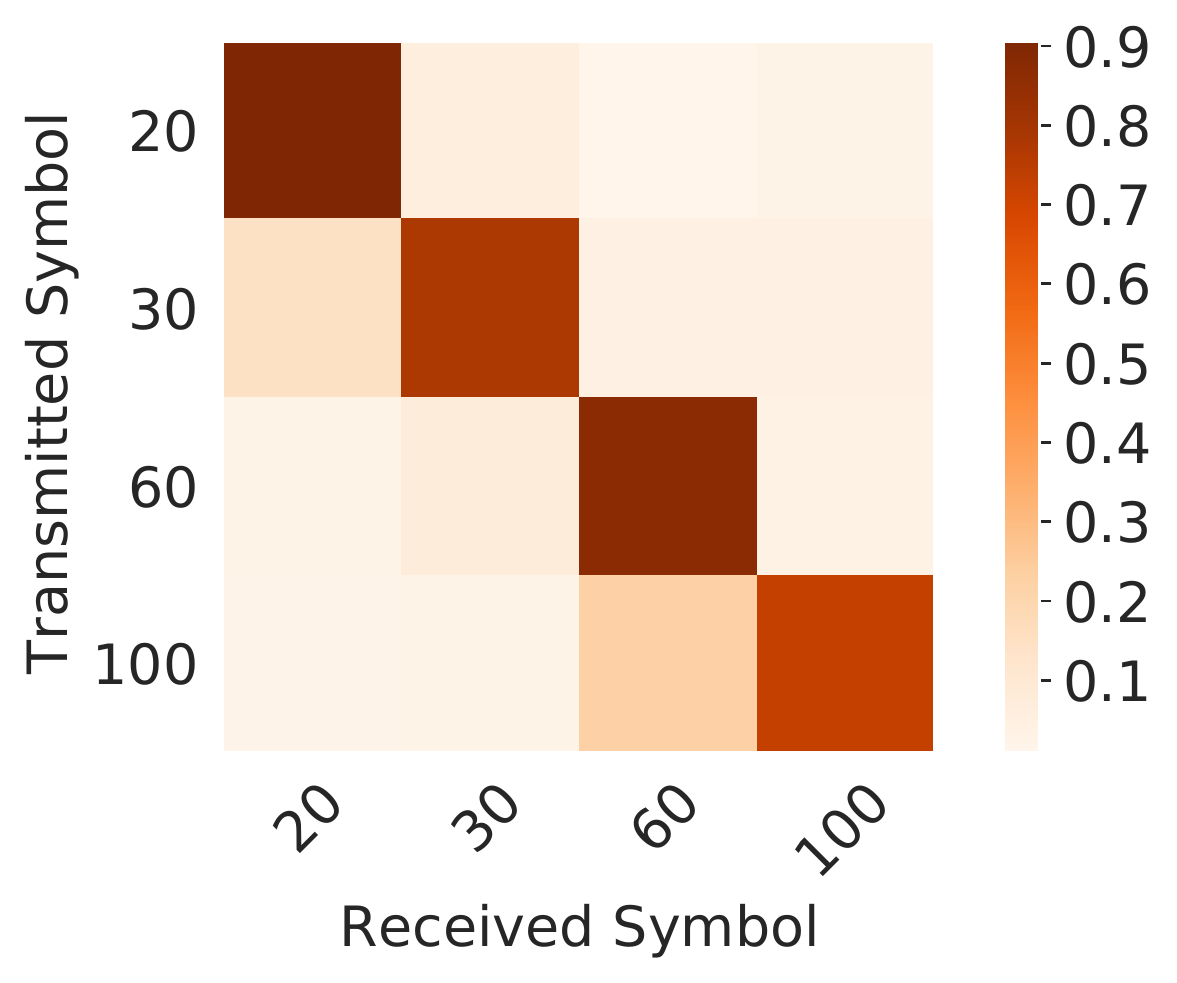}
    \caption{}
    \label{conf_m}
\end{subfigure}%
\hfill%
\begin{subfigure}[b]{0.48\linewidth}
    \includegraphics[width=\linewidth]{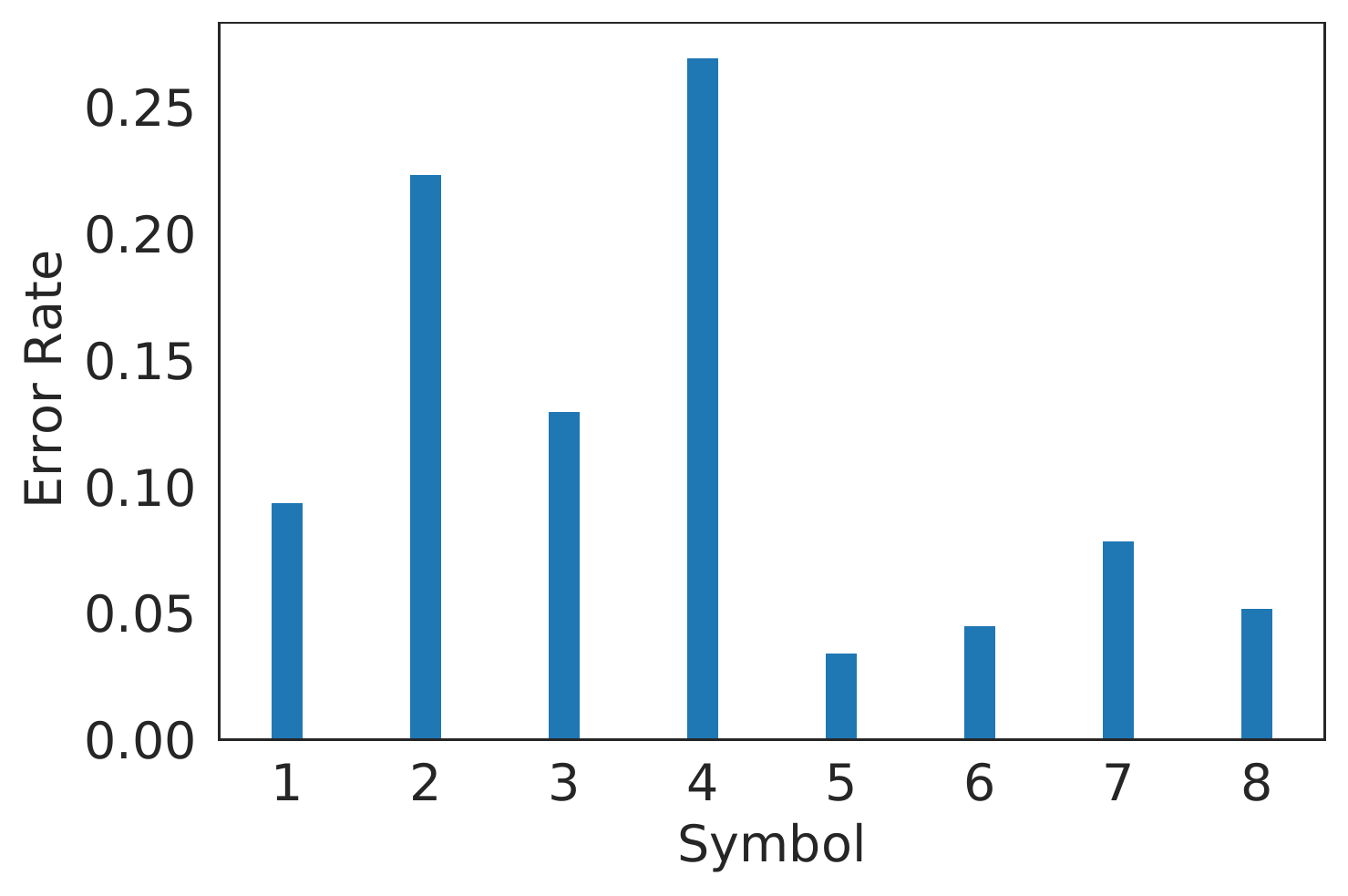}
    \caption{}
    \label{symbol_ber}
\end{subfigure} 
\caption{(a) Confusion matrix of transmitted and received symbols. (b) Per symbol error rate.}
\end{figure}

We also analyze each of the individual parameter modes in our algorithm which we use to encode data, i.e. PWMs, \texttt{ON} times and \texttt{OFF} times. We analyze the error rates of these modes individually to understand which modes work the best, in terms of bit rates or bit error rates. \cref{mode_overall_er} shows individual error rates for each of these modes. We see that the time-based modes (\texttt{ON} and \texttt{OFF}) perform the best with overall error rates less than 0.02, while the frequency-based mode (PWMs) show a relatively higher error (0.10). Although the time-based modes work more reliably than the PWM, the amount of modulation that can be done using time-based modes would be limited due to them being directly affecting the bit rates/transmission speeds. \cref{symbol_ber} indicates the error rates for each individual symbol where numbers 1-4 are PWMs, 5 and 6 are \texttt{ON} times and 7 and 8 \texttt{OFF} times. This further shows that time-based symbols (5-8) perform the best as opposed to the PWM-based symbols. The observations made in \cref{conf_m} is further clarified here as it can be seen that PWMs 30 and 100 show the highest error rates due to them being misidentified as 20 and 60, respectively.

\subsection{Acoustic Side-Channel} %
\label{sub:acoustic_side_channel}

Roy et al. \cite{roy2015ripple} recognized the possibility of information leakage via the noise produced during vibrations, and proposed a mechanism to jam the acoustic side-channel by emitting a noise from the transmitter 
to suppress the sound produced by vibrations. They further studied the \emph{sound of vibrations (SoV)} for different surfaces that the transmitter may be placed upon, and observed that glass surfaces cause the highest side-channel leaks, i.e. produces the loudest noise. To understand the severity of such a vulnerability in our proposed system, we similarly measure the SoV when a hand is used as the communication medium. %
\Cref{fig:SoV} provides the ratio of SoV of our custom and consumer-grade device setup (recorded 2 \emph{ft} away from the transmitter) to the ambient noise at various locations including, a school laboratory, an apartment, inside a moving vehicle and in a supermarket.
The lower the ratio (closer to zero), the quieter the SoV is relative to the ambient noise, thus, reducing the risk of acoustic leakage.
We observe the SoV to ambient noise ratio to be high in quieter locations such as an apartment or a school lab, while the ratio falls below 1 for louder locations such as a vehicle or a supermarket. This is in contrast to Roy et al.'s \cite{roy2015ripple} results, where the ratio was around 1.5 for all these locations. 
Compared to the consumer-grade smartphone setup, our custom setup (as seen in \cref{fig:SoV}) produces much louder sound ratios, which is likely because of the vibration motor being mounted on the custom handheld device without an enclosure.
The acoustic leakage from \name is significantly lower when consumer-grade mobile devices are used due to the enclosed nature of the vibration motor in such devices. 
We believe that this threat can be minimized by employing an acoustic jamming mechanism similar to the one proposed by Roy et al. \cite{roy2015ripple} by utilizing a speaker near the transmitting device.

\begin{table}[]
\centering
\caption{Sound of vibrations to ambient noise ratio.} 
\label{fig:SoV}
\begin{adjustbox}{width=0.99\linewidth}
\begin{tabular}{lllll}
\toprule
Setup & Lab & Apartment & Vehicle & Supermarket \\ 
\hline
Custom hardware setup     & 2.11 &   2.20  & 0.92  &     0.80\\ 
Consumer level smartphone & 1.46 &   1.52  & 0.63  &     0.58 \\
\bottomrule
\end{tabular}
\end{adjustbox}
\end{table}

\subsection{Practical Significance of the Bit Rate} %
\label{sub:practical_significance_of_the_bit_rate}

To understand the practical usefulness of the bitrate afforded by \name, we further analyze the transfer time for practical device-to-device secrets such as 4-digit PINs and 8-character passwords. For this analysis, we consider the data transfer scenario from the hand-held device to the wrist-wearable at a distance of 7.5 $cm$.
We observe that transfer of a 4-digit PIN takes a little more than 5 seconds, while an 8-character password takes approximately 10 seconds. Although these transfer speeds are considerably slower compared to short-range radio technologies such as Bluetooth and NFC, it must be noted that \name is envisioned to be only used as a potential secure side-channel to supplement traditional radio-based communication channels. Specifically as observed above, \name can enable usage scenarios such as securely proving device co-location or secure authentication in the absence of reliable radio channels by enabling sharing of short secrets at reasonable side-channel speeds.

\subsection{Energy Requirements} %
\label{sub:battery_consumption}
To evaluate the energy requirements of \name, we conducted an experiment to measure the amount of battery energy consumed over time. For the transmitter, we tested on two smartphones by sending 30 messages over a period of 1 hour. For the receiver, we tested two smartwatches by running the motion sensor data collection for 30 messages over a period of 1 hour. 
As seen in \cref{tab:BatCon}, energy consumption is only around 150 $mAh$ for Moto G7 Plus (2019) which is a newer device with Android 10, compared to the 300 $mAh$ for an older smartphone, Nokia 6 (2016), with Android 9. The receiver smartwatches show a similar energy consumption pattern with the newer Ticwatch with only 12 $mAh$ consumption compared to 14 $mAh$ in the older Sony W3. 

\begin{table}[h]
\centering
\caption{Energy consumption.} 
\label{tab:BatCon}
\begin{adjustbox}{width=0.8\linewidth}
\begin{tabular}{lll} 
\toprule
Device & Battery Capacity & Energy Consumed \\ 
\hline
Sony W3 & 420 $mAh$ & 14 $mAh$ \\
Mobovi Ticwatch E & 300 $mAh$ & 12 $mAh$ \\
Nokia 6 & 3000 $mAh$  & 300 $mAh$ \\
Moto G7 Plus & 3000 $mAh$ & 150 $mAh$ \\
\bottomrule
\end{tabular}
\end{adjustbox}
\end{table}

\subsection{Comparison with Previous Works}

\begin{table*}[t]
\centering
\caption{Comparison with related works.}
\begin{adjustbox}{width=0.99\linewidth}
\begin{tabular}{llllll} 
\toprule
Related Work & Setting & Sensors & Technique Used & Data Rate & Performance \\ 
\hline
VibeRing \cite{sen2020vibering} & Ring on finger to device & Accelerometer & Machine Learning based & 12.5 $bps$ & 0.05 (BER) \\
Ripple \cite{roy2015ripple} & 2 devices on a solid surface & Accelerometer & Frequency domain based  & 200 $bps$ & 0.017 (BER) \\
Vib-Connect \cite{yonezawa2011vib} & 2 devices in direct contact (mobile phone and laptop) & Accelerometer & Time domain thresholding based & Not mentioned & 100\% (Accuracy) \\
SYNCVIBE\cite{lee2018syncvibe} & 2 devices in direct contact & Accelerometer & Time domain based  & 20 $bps$ & 0.005 (BER) \\
SecureVibe\cite{kim2015vibration} & Device on top of a implanted medical device & Accelerometer & Time domain based  & 20 $bps$ & Not mentioned \\
Skin Sense (this work) & Between handheld device to wrist-wearable via skin & Accelerometer & Frequency \& time domain domain based & 6.6 $bps$ & 0.10 (BER) \\
\bottomrule
\end{tabular}
\end{adjustbox}
\end{table*}

Sen and Kotz \cite{sen2020vibering} proposed a vibration-based communication scheme using a smart ring, where the smart ring act as the vibration transmitter to communicate with Internet-of-Things (IoT) devices embedded with accelerometers. They were able to achieve a \emph{BER} of 0.05 with a bandwidth of 12.5 $bps$. Similar to many of the above mentioned works, their smart ring transmitter was required to be \emph{in direct contact} with the receiver IoT device for reliable communication.
The closest work to ours, in terms of using the human body/skin as the communication medium, is by Ma et al. \cite{ma2020skin}. They proposed a Multiple-Input-Multiple-Output (MIMO) communication scheme using two vibration motors and two piezo transducers. 
Due to limitations of vibration motors such as ramping time and the volatile nature of human skin as a channel, they use two motion sensors (acceleromter+gyroscope) embedded at the transmitter side, in order to acquire and utilize additional Channel State Information (CSI) by employing deep learning. 
Their MIMO scheme was able to achieve a MIMO capacity of about 5 $bps/Hz$, which is more than twice the capacity that could be achieved using a comparable Single-Input-Single-Output setup. However, there are several drawbacks in their proposal. First, their scheme is not very practical because it relies on extremely customized hardware and setup, often not found in commercial mobile and IoT devices.
Moreover, their scheme employs deep learning algorithms that typically require large amounts of training data for acceptable performance, which may not be trivial to obtain for all communication settings, conditions, and individual users. Lastly, performance evaluation of their scheme was done using data from only two human subject participants, and so it is unclear how generalizable their results are.
Another similar work (SecureVibe) which uses skin as a medium was proposed by Kim et al. \cite{kim2015vibration} using a smartphone vibration motor and a custom made implantable medical device equipped with an accelerometer. Although, their work achieves bit rates around 20 $bps$, their setup requires the smartphone to be directly on top of the implanted device under the skin with only 1 $cm$ distance between the motion sensor and the smartphone. We demonstrate that \name framework could be effective up to a distance of 15 $cm$ between the handheld smartphone (vibration motor) and the wrist wearable (motion sensor). Further, their work was only evaluated using an emulated human body model and not on actual human subject participants.
In contrast, \name is proposed for two externally held/worn devices, and the communication protocol was tested with human subject participants.
Shah et al. specifically studied the vibration propagation via skin of human arm and reported that even at a mere distance of 4 $cm$ \cite{shah2019vibration}, the vibration intensity drops around 70-80\% and continues to drop over 90\% at distances over 8 $cm$. 
In contrast, in a rigid medium such as a wooden board, the authors of \emph{Ripple} \cite{roy2015ripple}, observe that vibration intensity gradually increases up to 15 $cm$ before attenuating. This allows them to use 10 different vibration amplitudes to transmit data and still be able to accurately distinguish them during the demodulation to achieve a relatively higher data rate of 200 $bps$. It should be noted that on a medium with low rigidity such as the human skin, propagation of the vibration signal attenuates much more quickly at longer distances. This makes \name's design challenge a rather non-trivial one and with the same token makes adaptation of techniques employed by protocols such as \emph{Ripple} \cite{roy2015ripple} in this scenario infeasible.
}

%% file: discussion.tex
\section{Discussion} %
\label{sec:discussion}

\textbf{User's Perception and Preferences.}  
As vibrations carried by the human skin/body is perceptible to the end-user, we deployed a short survey to our 13 participants to gauge their feelings about the \name protocol. 
Based on the received survey responses, 38\% of the participants noted that they were not bothered by \name's operation, while 54\% were bothered slightly and only 8\% were highly bothered.
Although users can take advantage of \name by easily switching (or transfering) the hand-held device to the wrist-wearable device hand, we also studied users' preferences regarding \name usage.
When asked about which hand they use to hold a handheld device/smartphone, around 62\% answered that they use the same hand as the wrist wearable wearing hand followed by 23\% who use both the hands. The remaining 15\% said that they use their non-smartphone holding hand to wear the wrist wearable. Findings of this short survey does indicate that a vibration-based communication protocol such as \name is usable (from an end-user perspective) in practice, although more extensive usability and user-satisfaction surveys are needed.

\textbf{Other Side-channels.}  
In addition to the acoustic side-channel threat, there is a possibility of an attacker physically attaching an eavesdropping motion sensor to the communication surface \cite{roy2015ripple}. 
We believe that such a type of threat is unlikely in our proposed communication setup/protocol because we use the human body/skin as the communication channel. 
An adversary is unlikely to be able to directly attach an eavesdropping device to a victim user's body/skin without their cognizance. The transmitted messages could also be encrypted to further minimize the possibility of a contact based attack.
Further from the security perspective, as \name's main goal is to provide a secure channel against external eavesdropping devices, we assume that both the transmitter and receiver devices are fully trusted, executing only trusted (or non-malicious) apps.

%% file: related.tex
\section{Related Works} %
\label{sec:related_works}

Vibration-based communication techniques that employ vibration motors and motion sensors (esp. accelerometers) have been previously studied for various forms of underlying communication mediums (or channels), such as hard surfaces, direct device-to-device contact, and also via human skin.
Yonezawa et al. \cite{yonezawa2011transferring,yonezawa2011vib} proposed a mechanism to send information from a smartphone to a laptop computer and achieved a data rate up to 10 $bps$. Their proposal encodes information in a vibration signal emitted by the smartphone, which needs to be kept in \emph{physical contact} of the laptop, and the laptop detects the vibrations (with the encoded information) via an embedded or on-board accelerometer. 
Lee et al. \cite{lee2018syncvibe} proposed a similar communication framework (SYNCVIBE) where the 2 devices require \emph{physical contact} using a smartphone as a vibration transmitter and an external accelerometer device as a receiver attached to it and were able to achieve data rates around 20 $bps$ with a BER of 0.005.
Hwang et al. \cite{hwang2012privacy} proposed a similar communication mechanism, but between two smartphones placed on a solid surface, such as a wooden table, metal or plastic shelf, a stack of paper, and a cushioned chair. 
Their scheme achieved over 90\% accuracy for all the surfaces when the devices are placed \emph{roughly 25 $cm$} from each other. However, the accuracy drops significantly when the two devices are too close (10 $cm$) or too far apart.

Roy et al. \cite{roy2015ripple} proposed a similar communication framework, named \emph{Ripple}, which achieved a data rate of up to 200 $bps$ by using custom off-the-shelf vibration motors and accelerometer chips, and up to 80 $bps$ by using consumer smartphones. However, like Hwang et al. \cite{hwang2012privacy}, they employed \emph{solid surfaces} such as wooden and glass tables as the communication medium (or channel) in their scheme.
In a follow-up effort, Roy et al. \cite{roy2016ripple} proposed another vibration-based communication technique, but this time by using a microphone instead of an accelerometer as the receiver. Evaluation of their follow-up proposal showed that a smart ring with a vibration motor can achieve a bandwidth or data rate of around 7 $kbps$, while for a smartwatch the bandwidth drops to 2 $kbps$. 
However, an important requirement of their scheme was that the transmitting device had to be in \emph{close proximity} to the microphone-based receiver to achieve these bandwidths. 

In addition to these wearable device based related works, several other works have explored cyberphysical systems such as UAVs to propose the use of acoustics of their motors for communication and fingerprinting \cite{bannis2020bleep, ramesh2019sounduav}.

%% file: conclusion.tex
\section{Conclusion} %
\label{sec:conclusion}

In this paper we explored a novel form of communication between a handheld device and a wrist wearable by using a vibration motor transmitter and accelerometer-based receiver with human skin/hand being used as the communication medium. Since the human hand could have various anatomical and biomechanical differences among different people, we tested our proposed scheme under multiple realistic settings with 13 human subjects. Our proposed scheme was able to achieve a sustainable bandwidth of 6.66 \emph{bps} while keeping the BER below 0.10. %
Although, the current consumer level smartphones and wrist wearables have limitations, resulting in our scheme not being able to perform optimally, we believe that our work opens up further research in the area related to vibrations and human body-area communication channels.